\documentclass[mathpazo]{cicp}
\usepackage{amsmath}
\usepackage{graphicx}
\usepackage{alltt}

\begin{document}
\title{A Discontinuous Galerkin Method for Ideal Two-Fluid Plasma Equations}
%\author[J. Loverich and A. Hakim and U. Shumlak]{Tech-X Corporation\corrauth, Tech-X Corporation, University of Washington} 
%\author[A. Hakim]{Tech-X Corporation}
\author[J. Loverich et.~al.]{John Loverich\affil{1}\comma\corrauth, Ammar Hakim\affil{1} and Uri Shumlak\affil{2}}
%\address{\afillnum{1}\ TechX Corporation. \affilnum{2} University of Washington}
%\emails{{\tt loverich@txcorp.com}(J.~Loverich),{\tt ammar@txcorp.com}(A.~Hakim)},{\tt shumlak@aa.washington.edu}(U.~Shumlak)}
%\author[A. Hakim]{Ammar Hakim}
%\address{Tech-X Corporation}
%\author{U. Shumlak}{University of Washington}\thanks{Supported by AFOSR Grant No. F49620-02-1-0129}
%\corauth[cor]{Corresponding author.  Address: loverich@txcorp.com}
%\ead{jlovric@u.washington.edu}
%\author{A. Hakim},
%\author{U. Shumlak}
%\address{University of Washington, Aerospace and Energetics Research Program, Seattle, WA 98195-2250}
%\thanks{Supported by AFOSR Grant No. F49620-02-1-0129.  Simulations performed on the Ladon cluster in the department of Mechanical Engineering 
%at the University of Washington.}

\begin{abstract}
A discontinuous Galerkin method for the ideal 5 moment two-fluid plasma system is presented.  The method uses a second
or third order discontinuous Galerkin spatial discretization and a third order TVD Runge-Kutta time stepping
scheme.  The method is benchmarked against an analytic solution of a dispersive electron acoustic square
pulse as well as the two-fluid electromagnetic shock \cite{Shumlak2003} and existing numerical solutions to the 
GEM challenge magnetic reconnection problem \cite{Birn2001}.  The algorithm can be generalized to arbitrary 
geometries and three dimensions.  An approach to maintaining small gauge errors based on error propagation is suggested.
\end{abstract}

\maketitle

\section{Introduction}
Fusion power promises to be a safe, efficient and environmentally friendly energy source.  Controlled
fusion power concepts have been under investigation for decades, the vast majority of these concepts require
an intimate understanding of plasma physics to determine the stability and confinement properties.  Numerical
plasma physics has proved extremely valuable in deciphering experimental data and predicting the behavior of
plasma experiments.  Many plasma fluid models, and in particular the full two-fluid plasma model, have received very
little attention from the numerical plasma physics community.  This work describes an advanced algorithm
for the ideal 5-moment two-fluid plasma system.

To solve problems in plasma physics and to gain physical intuition of plasma phenomena a hierarchy of classical
plasma models have been developed.  The most fundamental continuum plasma model is the Vlasov model which eliminates individual particles in favor
of a continuous distribution function.  This model is six dimensional
as the distribution function is a function of both position and velocity.  The Vlasov model can be re-written as an equivalent
system that consists of an infinite number of moment equations.  A reduction of the Vlasov model can
then be obtained by truncating this infinite series.  Assuming scalar pressure and setting the heat tensor and higher moments to zero
produces the 5 moment truncation of the Vlasov model.  This model is known as the ideal 5 moment two-fluid plasma model, and will be
discussed in this paper.  Asymptotic approximations of this two-fluid system produce a series of increasingly 
simpler fluid models including two-fluid MHD (Magnetohydrodynamics), Hall MHD and then the ideal MHD models.

The main benefit of a fluid model over the Vlasov model is the reduced dimensionality from 6 dimensions to 3 dimensions.
Physics is lost in this reduction, but an enormous amount of physics relevant to fusion and spacecraft propulsion
remains in the fluid description.  Ideal MHD has been extremely successful in explaining large scale instabilities in
such devices as the Z-pinch, spheromak and tokamak \cite{Freidberg87,Bellan2000}.  Unfortunately there are many regimes where the description is
invalid and where it fails to explain the observed phenomena.  An example of this includes ion demagnetization which is
important in Field Reversed Configurations \cite{Hakim2007} and Hall thrusters.  Hall
MHD addresses both these issues but fails to describe other plasma phenomena such as the demagnetization of electrons
in regions of low magnetic field which is important in collisionless reconnection.  The two-fluid MHD approach adds terms such as electron 
inertia which is an important mechanism for breaking 
the frozen in flux condition for electrons as it acts as a ``dissipation" mechanism in the absence of resistivity \cite{Connor93}.  
The quasi-neutrality condition still constrains the electron and ion motions, to allow complete independence of electron and ion 
motion the quasi-neutrality condition must be relaxed; the result is the ideal two-fluid plasma system.

Two-fluid effects are important in the generation of turbulence through microinstabilities.
Most plasmas are turbulent at some scale, however the simplest fluid model, ideal MHD, describes plasmas physics that is more or 
less laminar where the two-fluid model produces turbulent phenomena.  This can be explained in part by the fluid description of electrons. 
In a two-fluid model both the electrons and the ions may become unstable independently.  In particular, electrons carry most of the 
current in an MHD plasma.  This current may produce a large amount of differential motion in the electron fluid when magnetic field 
gradients are present even if the plasma is in a static MHD equilibrium.  The generation of microturbulence through processes such as
the lower hybrid drift instability and the modified two-stream instability may be important in both Z-pinch and theta-pinch plasmas.
These instabilities are frequently cited as sources of anomalous resistivity \cite{Krall71}, magnetic diffusion and heating and in certain cases may ultimately 
drive macroscopic MHD instabilities \cite{Davidson75}.  

A particularly good application of the
two-fluid plasma model is the fusion Z-pinch \cite{Loverich2006}.  Many plasma experiments last a few seconds whereas the shortest plasma 
times scale, the electron plasma oscillation, can occur on the scale of pico seconds.  However, in the case of the fusion
Z-pinch these time scales can be compressed to about 4 orders of magnitude between the shortest time scale,
the electron plasma period the MHD instability growth time, which puts two-fluid Z-pinch simulations in the range 
of numerical methods.
Conceptual Z-pinch fusion reactors are high density with extremely strong magnetic fields which makes the two-fluid plasma
system particularly applicable \cite{Solovev84}.  Furthermore the radius of the pinch can be ~1000 Debye lengths in 
some designs \cite{Haines82} which
means Debye length scales could be resolved in very high resolution simulations.  
Artificially increasing the electron mass to ion mass ratio and increasing the ratio of the Alfven speed to the speed of light 
can make the Z-pinch problem more computationally tractable while maintaining the relevant physics.  
The analysis of microinstabilities such as the lower hybrid drift instability 
may be important in progressing towards a better understanding of Z-pinch physics.

Algorithms have been designed for various fluid plasma models including MHD \cite{Eberhardt96,Sovinec2004}, 
Hall MHD \cite{Bhattacharjee2003,Breslau2003,Huba2003,Park99} various forms of electrostatic two-fluid plasma
models \cite{Baboolal2001,Munz95,Rambo91} and the ideal two-fluid system \cite{Mason87,Mason86,Shumlak2003}.  The first well described two-fluid 
plasma algorithm was the ANTHEM \cite{Mason87,Mason86} code.  It was used to simulate fast phenomena in high density plasmas inaccessible
to PIC codes, the applications included simulations of plasma opening switches and the fast igniter concept.  ANTHEM used a flux corrected transport 
(FCT) type algorithm for the fluids and an FDTD type algorithm for the fields.  This type of algorithm is difficult to extend to general geometries
because of the staggered scheme used for Maxwell's equations.
In \cite{Shumlak2003} a full two-fluid algorithm using the finite volume method for both fluids and fields was described for one-dimensional problems. 
An issue with the finite volume algorithm was the decay of equilibrium solutions due to the 
low order of accuracy which resulted from the source term integration and diffusive limiters used.  Major improvements on the finite volume technique
have been made with the help of various divergence cleaning techniques and careful attention to source term treatment.  An improved finite volume
approach is described in \cite{Hakim2006}.  

The purpose of the paper is to develop a numerical algorithm for the ideal 5 moment two-fluid plasma 
system using the discontinuous Galerkin method so that it can be easily generalized to arbitrary geometries 
and to arbitrarily high order accuracy to help capture plasma instabilities.
TVB discontinuous Galerkin methods are described in \cite{Cockburn89,Cockburn89b,Cockburn90}.  They are extended to the multi-dimensional Euler equations
in \cite{Cockburn98} and to Maxwell's equations in \cite{Hesthaven2002,Cockburn2004}.  The two-fluid system of equations consists of two sets of 
Euler equations, one for the electrons and the other for the ions, and the complete Maxwell's equations.  The ideal two-fluid system differs from 
the ideal MHD equations in that it is composed of three separate (but well understood) hyperbolic systems coupled through source terms.  The MHD
equations are, on the other hand, a unique hyperbolic system.  A discontinuous Galerkin method for the MHD equations was developed in \cite{Warburton99} and for two temperature MHD in \cite{Lin2006}.  The technique is used in a Vlasov-Maxwell algorithm in \cite{Mangeney2002}, numerous other applications can be found in \cite{Cockburn2000}.

In section \ref{S:TwoFluidModel} the ideal 5-moment two-fluid model is described and the equations
are presented.  In section \ref{S:ScalarProblem} a scalar model problem is derived from the two-fluid systems which helps to illustrate
some of the numerical issues with the system.  In section \ref{S:DiscontinuousGalerkinMethod} the discontinuous 
Galerkin method applied to the two-fluid plasma system is presented.  In section \ref{S:Simulations} simulations are presented including
an electron acoustic pulse for validation of code accuracy, an electrostatic shock, the two-fluid electromagnetic shock \cite{Shumlak2003}, 
and the GEM challenge magnetic reconnection problem \cite{Birn2001} where the reconnected magnetic flux can be
compared to published results.  Finally, in section \ref{S:Discussion} the conclusions are discussed.

\section{Two-Fluid Model}\label{S:TwoFluidModel}
The full two-fluid plasma model consists of a set of fluid equations for the electrons and ions
plus the complete Maxwell's equations including displacement current.  The fluid and electromagnetic systems are coupled
by Lorentz forces and current sources.  In the following equations, $\mathbf{E}$ is
the electric field, $\mathbf{B}$ is the magnetic field, $q_{s}$ is the species charge (subscript $s$ is $i$ for ions and $e$ for electrons), 
$\rho_{s}$ is the species density, 
$m_{s}$ is the species mass, $\mathbf{U}_{s}$ is the species velocity, $P_{s}$ is the species pressure and $e_{s}$ is the species total energy with 
$e_{s}=\frac{1}{2}\rho_{s}\,\mathbf{U}_{s}^{2}+\frac{1}{1-\gamma_{s}}p_{s}$.  
The species number density is defined as $n_{s}=\frac{\rho_{s}}{m_{s}}$, $\epsilon_{0}$ is the permittivity and $\mu_{0}$ is the
permeability of free space.  Maxwell's equations are presented in SI units.  The complete Ampere's law
is used
\begin{equation}\label{E:Ampere}
	\partial_{t}\,\mathbf{E}-c^{2}\left(\nabla\times\,\mathbf{B}\right)=-\frac{1}{\epsilon_{0}}\sum_{s}\frac{q_{s}}{m_{s}}\rho_{s}\mathbf{U}_{s}\,,
\end{equation}
and the complete Faraday's law
\begin{equation}\label{E:Faraday}
	\partial_{t}\,\mathbf{B}+\left(\nabla\times\,\mathbf{E}\right)=0\,.
\end{equation}
The magnetic flux equation,
\begin{equation}\label{E:MagneticFlux}
	\nabla\cdot \mathbf{B}=0
\end{equation} 
and Poisson's equation\thinspace\eqref{E:Poisson},
\begin{equation}\label{E:Poisson}
	\nabla\cdot \mathbf{E}=\frac{1}{\epsilon_{0}}\left(q_{i}\,n_{i}+q_{e}\,n_{e}\right)
\end{equation}
 are constraint equations
which can be derived from Ampere's law\thinspace\eqref{E:Ampere}, Faraday's law\thinspace\eqref{E:Faraday} as well as the species
continuity equation given below in\thinspace\eqref{E:SpeciesContinuity} under the assumption that the constraints are satisfied initially.  
A simple way to reduce the error in the constraint equations is to use
the perfectly hyperbolic Maxwell's equations where equations\thinspace\eqref{E:Ampere}, \eqref{E:Faraday}, \eqref{E:MagneticFlux}, \eqref{E:Poisson}
are modified, so that Ampere's law becomes
\begin{equation}\label{E:Ampere2}
	\partial_{t}\,\mathbf{E}-c^{2}\left(\nabla\times\,\mathbf{B}\right)+\nabla\psi_{E}=-\frac{1}{\epsilon_{0}}\sum_{s}\frac{q_{s}}{m_{s}}\rho_{s}\mathbf{U}_{s}\,,
\end{equation}
Faraday's law becomes
\begin{equation}\label{E:Faraday2}
	\partial_{t}\,\mathbf{B}+\left(\nabla\times\,\mathbf{E}\right)+\nabla\psi_{B}=0\,.
\end{equation}
The magnetic flux equation becomes,
\begin{equation}\label{E:MagneticFlux2}
	\frac{1}{\Gamma_{B}^{2}}\frac{\partial \psi_{B}}{\partial t}+\nabla\cdot \mathbf{B}=0
\end{equation} 
and Poisson's equation becomes,
\begin{equation}\label{E:Poisson2}
	\frac{1}{\Gamma_{E}^{2}}\frac{\partial \psi_{E}}{\partial t}+\nabla\cdot \mathbf{E}=\frac{1}{\epsilon_{0}}\left(q_{i}\,n_{i}+q_{e}\,n_{e}\right)\,.
\end{equation}
Where $\psi_{B}$ and $\psi_{E}$ are correction potentials and $\Gamma_{B}$, $\Gamma_{E}$ are the correction potential wave speeds
which propagate errors in the divergence constraints out at the speeds $\Gamma_{B}$ and $\Gamma_{E}$.  The correction potential wave speeds
can be set to zero for problems where the correction terms are not necessary.  Details of the correction potential technique
are described in \cite{Munz2000} and used in our previous work \cite{Hakim2006}.  In many problems of experimental interest electrons flow from a surface into
the simulation domain.  In these situations we believe more sophisticate divergence preservation will be required.  It's
important to maintain exact charge conservation especially when the emission is space charge limited.  For this type of problem a constrained
transport approach should be taken.  Charge conserving constrained transport is frequently used 
in particle in cell (PIC) codes \cite{Umeda2003}, \cite{Villasenor1992}, \cite{Mardahl1997}.  In addition, $\nabla\cdot B=0$ preserving constrained transport
is frequently used in the MHD system \cite{Li2008}, \cite{Balsara2009}, \cite{Gardiner2008}, \cite{Londrillo2004}.
 
The fluid equations are simply the inviscid Navier Stokes equations with Lorentz force source terms.  Each fluid species
has its own equation for energy,
\begin{equation}\label{E:SpeciesEnergy}
	\partial_{t}\,e_{s}+\nabla\cdot\left(\mathbf{U}_{s}\left(e_{s}+P_{s}\right)\right)=\frac{q_{s}}{m_{s}}\rho_{s}\mathbf{E}\cdot \mathbf{U}_{s}\,,
\end{equation}
momentum,
\begin{equation}\label{E:SpeciesMomentum}
	\partial_{t}\,\left(\rho_{s}\mathbf{U}_{s}\right)+\nabla_{\alpha}\left(\rho_{s}\mathbf{U}_{s}^{\alpha}\mathbf{U}_{s}\right)
	+\nabla_{\alpha}\left(\delta^{\alpha\,\beta}P_{s}\right)=\frac{q_{s}}{m_{s}}\,\rho_{s}\left(\mathbf{E}+\mathbf{U}_{s}\times\,\mathbf{B}\right)\,,
\end{equation} 
and continuity,
\begin{equation}\label{E:SpeciesContinuity}
		\partial_{t}\,\rho_{s}+\nabla\cdot\left(\rho_{s}\mathbf{U}_{s}\right)=0\,.
\end{equation}
This means that each species has its own
temperature, velocity and number density.  As a result, quasi-neutrality is not assumed and things like electron plasma waves and
 ion subshocks should be observed numerically.  This system is identical to the system used 
in \cite{Shumlak2003}. 

The ideal two fluid plasma system can be written as three systems of balance laws,
\begin{equation}\label{E:ElectronBalanceLaw}
	\frac{\partial\,Q_{e}}{\partial t}+\nabla\cdot F_{e}\left(Q_{e}\right)=\psi_{e}\left(Q_{e},\,Q_{em}\right)\,,
\end{equation}
for the electron equations,
\begin{equation}\label{E:IonBalanceLaw}
	\frac{\partial\,Q_{i}}{\partial t}+\nabla\cdot F_{i}\left(Q_{i}\right)=\psi_{i}\left(Q_{i},\,Q_{em}\right)\,,
\end{equation}
for the ion equations, and
\begin{equation}\label{E:EMBalanceLaw}
	\frac{\partial\,Q_{em}}{\partial t}+\nabla\cdot F_{em}\left(Q_{em}\right)=\psi_{em}\left(Q_{i},\,Q_{e}\right),
\end{equation}
for Maxwell's equations.  These balance laws, Eqns.\eqref{E:ElectronBalanceLaw}-\eqref{E:EMBalanceLaw}, are given
in full form by,
\begin{multline}
	\frac{\partial}{\partial\,t}
	\left(
	\begin{array}{c}
	\rho_{s}\\
	\rho_{s}\,U_{x\,s}\\
	\rho_{s}\,U_{y\,s}\\
	\rho_{s}\,U_{z\,s}\\
	e_{s}
	\end{array}
	\right)+
	\nabla\cdot
	\left(
	\begin{array}{ccc}
	\rho_{s}\,U_{x\,s} & \rho_{s}\,U_{y\,s} & \rho_{s}\,U_{z\,s}\\
	\rho_{s}\,U_{x\,s}U_{x\,s}+P_{s} & \rho_{s}\,U_{x\,s}\,U_{y\,s} & \rho_{s}\,U_{x\,s}\,U_{z\,s}\\
	\rho_{s}\,U_{y\,s}U_{x\,s} & \rho_{s}\,U_{y\,s}\,U_{y\,s}+P_{s} & \rho_{s}\,U_{y\,s}\,U_{z\,s}\\
	\rho_{s}\,U_{z\,s}U_{x\,s} & \rho_{s}\,U_{z\,s}\,U_{y\,s} & \rho_{s}\,U_{z\,s}\,U_{z\,s}+P_{s}\\
	U_{x\,s}\left(e_{s}+P_{s}\right) & U_{y\,s}\left(e_{s}+P_{s}\right) & U_{z\,s}\left(e_{s}+P_{s}\right)\\
	\end{array}
	\right)=\\
	\left(
	\begin{array}{c}
	0\\
	q_{s}\,n_{s}\left(E_{x}+U_{y\,s}\,B_{z}-U_{z\,s}\,B_{y}\right)\\
	q_{s}\,n_{s}\left(E_{y}+U_{z\,s}\,B_{x}-U_{x\,s}\,B_{z}\right)\\
	q_{s}\,n_{s}\left(E_{z}+U_{x\,s}\,B_{y}-U_{y\,s}\,B_{x}\right)\\
	q_{s}\,n_{s}\left(E_{x}\,U_{x\,s}+E_{y}\,U_{y\,s}+E_{z}\,U_{z\,s}\right)\\
	\end{array}
	\right)
\end{multline}
\begin{multline}
	\frac{\partial}{\partial t}
	\left(
	\begin{array}{c}
	B_{x}\\
	B_{y}\\
	B_{z}\\
	E_{x}\\
	E_{y}\\
	E_{z}
	\end{array}
	\right)+
	\nabla\cdot\,\left(
	\begin{array}{ccc}
	0 & E_{z} & -E_{y}\\
	-E_{z} & 0 & E_{x}\\
	E_{y} & -E_{x} & 0\\
	0 & -c^{2}\,B_{z} & c^{2}\,B_{y}\\
	c^{2}\,B_{z} & 0 & -c^{2}\,B_{x}\\
	-c^{2}\,B_{y} & c^{2}\,B_{x} & 0\\
	\end{array}
	\right)=\\
	\left(
	\begin{array}{c}
	0\\
	0\\
	0\\
	-\frac{1}{\epsilon_{0}}\left(q_{e}\,n_{e}\,U_{x\,e}+q_{i}\,n_{i}\,U_{x\,i}\right)\\
	-\frac{1}{\epsilon_{0}}\left(q_{e}\,n_{e}\,U_{y\,e}+q_{i}\,n_{i}\,U_{y\,i}\right)\\
	-\frac{1}{\epsilon_{0}}\left(q_{e}\,n_{e}\,U_{z\,e}+q_{i}\,n_{i}\,U_{z\,i}\right)\\
	\end{array}
	\right)
\end{multline}

\section{Derivation of a Scalar Model Problem}\label{S:ScalarProblem}
The ideal 5-moment two-fluid system is unusual in that the source terms act
as harmonic oscillators, the source terms are purely dispersive without dissipation 
or amplification.  This fact is important when considering numerical methods to use, as
the source term integration should introduce as little dissipation as feasible in
order to avoid damping these oscillations.  It frequently occurs in a two-fluid plasma
that convective forces are in balance with oscillating sources to produce an equilibrium.
With this in mind, a simple model problem is derived which may help one choose a proper
numerical method for the two-fluid system. 

Linearizing the electron x-momentum equation\thinspace\eqref{E:SpeciesMomentum} and Ampere's law\thinspace\eqref{E:Ampere} while assuming a constant
background ion density and assuming that $\mathbf{B}=0$, 
a partial differential equation for electron plasma oscillations can be derived which takes the form
\begin{equation}\label{E:ElectronAcousticPDE}
	\frac{\partial^{2}u}{\partial\,t^{2}}=-\omega_{p\,e}^{2}u\,,
\end{equation}
where $u$ is the perturbed x velocity and $\omega_{p\,e}$ is the
electron plasma frequency.  This can be transformed to a first order equation in complex variables
by making the transformation $\omega_{pe}v=\frac{\partial u}{\partial t}$ and letting $Q=v+i\,u$.
\begin{equation}\label{E:SimpleHarmonicOscillator}
	\frac{\partial\,Q}{\partial\,t}=i\,\omega_{p\,e}Q\,.
\end{equation}
By transforming variables $Q\left(x,t\right)\to Q\left(\eta,t\right)$ where $\eta=x+a\,t$, Equation\eqref{E:SimpleHarmonicOscillator} becomes
an advection oscillation equation
\begin{equation}\label{E:ModelEquation}
	\frac{\partial Q}{\partial t}+a\frac{\partial Q}{\partial \eta}=i\,\omega_{p\,e}\,Q\,,
\end{equation}
where $a$ is the wave propagation speed and $\omega_{p\,e}$ is the oscillation frequency. Any algorithm that is
stable to the two-fluid system should be stable when applied to Eqn\eqref{E:ModelEquation}.
This equation has solutions $Q=A\,e^{i\left(k\,\eta-\omega\,t\right)}$ where $\omega=a\,k-\omega_{p\,e}$.
In particular\eqref{E:ModelEquation} admits a steady state solution $Q=A\,e^{i\,k\,\eta}$ on an infinite domain
where the source term is in balance with the flux.  
It is important to note that there are an infinite number of equilibria which differ by a continuous range of scalar
factors $A$.  This point is important when considering
numerical methods for this system since a numerical method with too much dissipation could conceivably move a steady state solution from one
equilibrium to another all the while moving towards a state where $A=0$.  This loss of amplitude is also observed in equilibrium type problems in the two-fluid
system.  An effective numerical algorithm must be both stable to the advection equation and oscillation equation and must have low dissipation
in equilibrium type problems.  The discontinuous Galerkin method is an ideal candidate for solving the two-fluid system.  Its accuracy can easily be increased to reduce numerical dissipation while being stable to both the advection equation and the oscillation equation.

\section{Runge-Kutta Discontinuous Galerkin Method}\label{S:DiscontinuousGalerkinMethod}
Discontinuous Galerkin methods are high order extensions of upwind schemes using a finite element
formulation where the elements are discontinuous at cell interfaces.  Details of the method
are discussed in \cite{Cockburn89,Cockburn89b,Cockburn90,Cockburn98} and reproduced here for our particular case.  
The balance law 
\begin{equation}
	\frac{\partial Q}{\partial t}+\nabla\cdot F\left(Q\right) = \psi\left(Q\right)
\end{equation}
is multiplied by the set of basis functions $\{v_{r}\}$ and integrated over the finite volume element $K$.  
For second order spatial accuracy the basis set on a unit square reference element is
\begin{equation}\label{E:SecondOrderBasis}
	\{v_{r}\}=\{v_{0},v_{x},v_{y}\}=\{1,x,y\}\,.
\end{equation}
For third order spatial accuracy 
\begin{equation}\label{E:ThirdOrderBasis}
	\{v_{r}\}=\{v_{0},v_{x},v_{y},v_{x\,y},v_{x\,x},v_{y\,y}\}=\{1,x,y,x\,y,x^{2}-\frac{1}{3},y^{2}-\frac{1}{3}\}
\end{equation} is used.  
The equation is written,
\begin{equation}
	\int_{K}\frac{\partial Q}{\partial t}\,v_{r}\,dV+\int_{K}\left(\nabla \cdot F\right)\,v_{r}\,dV = \int_{K}\psi\,v_{r}\,dV\,.
\end{equation}
Integrate by parts to get
\begin{equation}
	\int_{K}\frac{\partial Q}{\partial t}\,v_{r}\,dV+\int_{\partial\,K}\left(\tilde{F}\cdot n\right)\,v_{r}\,d\Gamma -
	\int_{K}F\cdot\left(\nabla\,v_{r}\right)\,dV= \int_{K}\psi\,v_{r}\,dV\,.
\end{equation}
The surface flux $\tilde{F}\cdot n$ is a numerical approximation of the exact flux $F\cdot n$ across the interface.
The discrete conserved variable $Q$ is defined as a linear combination of the basis functions inside an 
element $K$, with 
\begin{equation}
	Q=\sum_{r}\,v_{r}\,Q_{r}\,.
\end{equation}
The integral $\int_{K}\frac{\partial Q}{\partial t}\,v_{r}\,dV=\frac{\partial Q_{r}}{\partial t}\,C\,V$ where
$C$ is the constant $\frac{1}{V}\int_{K}\,v_{r}^{2}\,dV$ and V is the volume of the element.  Using these definitions we get the discrete equation
\begin{equation}
	\frac{\partial Q_{r}}{\partial t}C\,V+\sum_{e}\sum_{l}\,w_{l}\,\left(\tilde{F_{l}}\cdot n\right)\,v_{r\,l}\,\Gamma_{e} -
	\sum_{m}\,w_{m}\,F_{m}\cdot\left(\nabla\,v_{r\,m}\right)\,V= \sum_{m}\,w_{m}\,\psi_{m}\,v_{r\,m}\,V\,,
\end{equation}
when the integrals are replaced by appropriate Gaussian quadratures.  $\Gamma_{e}$ is the surface area of the cell face in consideration, 
$e$ refers to an element face, $l$ are quadrature points on a face with $w_{l}$ the associated weight.  $m$ refer to quadrature points in the volume with $w_{m}$ the
associated weight.  Functions with subscript $l$ or $m$ are evaluated at the $l^{th}$ face quadrature points and $m^{th}$ volume quadrature points
respectively. For a second order method
the edge integrals are replaced by a two point Gaussian quadrature 
\begin{equation}
	\int_{-1}^{1}f\left(x\right)dx \approx f\left(\frac{1}{\sqrt{3}}\right)+f\left(-\frac{1}{\sqrt{3}}\right)
\end{equation}
A four point quadrature is used for the volume integral given by,
\begin{multline}
	\int_{-1}^{1}\int_{-1}^{1}f\left(x,y\right)dx\,dy \approx f\left(\frac{1}{\sqrt{3}},\frac{1}{\sqrt{3}}\right)+f\left(-\frac{1}{\sqrt{3}},\frac{1}{\sqrt{3}}\right)+\\
		f\left(-\frac{1}{\sqrt{3}},-\frac{1}{\sqrt{3}}\right)+f\left(\frac{1}{\sqrt{3}},-\frac{1}{\sqrt{3}}\right)\,.
\end{multline}  
The discrete equations for the second order scheme using the basis functions given in\thinspace\eqref{E:SecondOrderBasis} become
\begin{subequations}\label{E:SecondOrderScheme}
	\begin{equation}
		\frac{\partial Q_{0}}{\partial t}V+\sum_{e}\sum_{l}\,w_{l}\,\left(\tilde{F_{l}}\cdot n_{e}\right)\,v_{0\,l}\,\Gamma_{e} = 
		\sum_{m}\,w_{m}\,\psi_{m}\,v_{0\,m}\,V\,,
	\end{equation}
	\begin{equation}
		\frac{\partial Q_{x}}{\partial t}V+3\sum_{e}\sum_{l}\,w_{l}\,\left(\tilde{F_{l}}\cdot n_{e}\right)\,v_{x\,l}\,\Gamma_{e} -
		3\sum_{m}\,w_{m}\,F_{m}\cdot\left(\nabla\,v_{x\,m}\right)\,V= 3\sum_{m}\,w_{m}\,\psi_{m}\,v_{x\,m}\,V\,,
	\end{equation}
	\begin{equation}
		\frac{\partial Q_{y}}{\partial t}V+3\sum_{e}\sum_{l}\,w_{l}\,\left(\tilde{F_{l}}\cdot n_{e}\right)\,v_{y\,m}\,\Gamma_{e} -
		3\sum_{m}\,w_{m}\,F_{m}\cdot\left(\nabla\,v_{y\,m}\right)\,V= 3\sum_{m}\,w_{m}\,\psi_{m}\,v_{y\,m}\,V\,.
	\end{equation}
\end{subequations}
The derivatives of the basis functions can be calculated analytically since the polynomial basis functions are known.  The discontinuous Galerkin
method is applied to each balance law\thinspace\eqref{E:ElectronBalanceLaw}\eqref{E:IonBalanceLaw}\eqref{E:EMBalanceLaw} at every time step.
For the third order space method the edge integrals are done using a 3 point quadrature 
\begin{equation}
	\int_{-1}^{1}f\left(x\right)dx \approx \frac{8}{9}f\left(0\right)+\frac{5}{9}\left(f\left(\frac{\sqrt{3}}{5}\right)+f\left(-\frac{\sqrt{3}}{5}\right)\right)
\end{equation}
The volume integrals are performed using a 9 point quadrature which can be calculated by doing a 3 point integration in the $x$ direction
and then a 3 point integration in the $y$ direction.  This produces the following approximate integral
\begin{multline}
	\int_{-1}^{1}\int_{-1}^{1}f\left(x,y\right)dx\,dy \approx \frac{64}{81}f\left(0,0\right)+\\
	\frac{25}{81}\left[f\left(\frac{\sqrt{3}}{5},\frac{\sqrt{3}}{5}\right)+
	f\left(-\frac{\sqrt{3}}{5},\frac{\sqrt{3}}{5}\right)+f\left(-\frac{\sqrt{3}}{5},-\frac{\sqrt{3}}{5}\right)+f\left(\frac{\sqrt{3}}{5},-\frac{\sqrt{3}}{5}\right)\right]+\\
	\frac{40}{81}\left[f\left(0,\frac{\sqrt{3}}{5}\right)+
	f\left(\frac{\sqrt{3}}{5},0\right)+f\left(0,-\frac{\sqrt{3}}{5}\right)+f\left(-\frac{\sqrt{3}}{5},0\right)\right]
\end{multline}
For the 3rd order scheme the following discrete equations must be updated in addition to those given by the second order scheme\thinspace\eqref{E:SecondOrderScheme} using the basis functions defined in\thinspace\eqref{E:ThirdOrderBasis} becomes,
\begin{subequations}\label{E:ThirdOrderScheme}
	\begin{equation}
		\frac{\partial Q_{xy}}{\partial t}V+9\sum_{e}\sum_{l}\,w_{l}\,\left(\tilde{F_{l}}\cdot n_{e}\right)\,v_{xy\,l}\,\Gamma_{e} -
		9\sum_{m}\,w_{m}\,F_{m}\cdot\left(\nabla\,v_{xy\,m}\right)\,V= 9\sum_{m}\,w_{m}\,\psi_{m}\,v_{xy\,l}\,V\,,
	\end{equation}
	\begin{equation}
		\frac{\partial Q_{xx}}{\partial t}V+\frac{45}{4}\sum_{e}\sum_{l}\,w_{l}\,\left(\tilde{F_{l}}\cdot n_{e}\right)\,v_{xx\,l}\,\Gamma_{e} -
		\frac{45}{4}\sum_{m}\,w_{m}\,F_{m}\cdot\left(\nabla\,v_{xx\,l}\right)\,V= \frac{45}{4}\sum_{m}\,w_{m}\,\psi_{m}\,v_{xx\,m}\,V\,,
	\end{equation}
	\begin{equation}
		\frac{\partial Q_{yy}}{\partial t}V+\frac{45}{4}\sum_{e}\sum_{l}\,w_{l}\,\left(\tilde{F_{l}}\cdot n_{e}\right)\,v_{yy\,l}\,\Gamma_{e} -
		\frac{45}{4}\sum_{m}\,w_{m}\,F_{m}\cdot\left(\nabla\,v_{yy\,m}\right)\,V= \frac{45}{4}\sum_{m}\,w_{m}\,\psi_{m}\,v_{yy\,m}\,V\,.
	\end{equation}
\end{subequations}
Though the spatial discretization uses a finite element approach, the time integration uses standard finite difference methods which are described in the
next section.  The algorithm described is an explicit finite element method, data is only exchanged between neighboring cells.  The solution does not
need to be continuous at cell interfaces which is particularly useful for problems with shocks.

\subsection{Time Integration Schemes}\label{S:TimeIntegration}
Time integration schemes that are stable for the advection equation must also be stable to the oscillation equation if they are
to be stable in general to the two-fluid system.  In this paper the 3rd order TVD Runge-Kutta method \cite{Cockburn89} is used,
\begin{subequations}
	\begin{equation}
		Q^{1}=Q^{n}+\Delta\,t\,L\left[Q^{n}\right]
	\end{equation}
	\begin{equation}
		Q^{2}=\frac{3}{4}Q^{n}+\frac{1}{4}\left(Q^{1}+\Delta\,t\,L\left[Q^{1}\right]\right)
	\end{equation}
	\begin{equation}
		Q^{n+1}=\frac{1}{3}Q^{n}+\frac{2}{3}\left(Q^{2}+\Delta\,t\,L\left[Q^{2}\right]\right)\,,
	\end{equation}
\end{subequations}
The time integration scheme is applied to each $Q_{r}$ at every time step to evolve the solution.  The term $L\left[Q^{n}\right]$
represents the entire ``left hand side" which is everything but the time derivative evaluated at $Q^{n}$.  
It is important to note that all two step second order Runge-Kutta schemes
are unstable to the oscillation equation \cite{Durran98}.
\subsection{Evaluating $\tilde{F}\cdot n$}
The flux $\tilde{F}\cdot n$ can be evaluated a number of different ways.  The local Lax flux is used in this paper and is computed
at each face as
\begin{equation}
	\tilde{F}\cdot n = \frac{1}{2}\left(F^{+}_{i}+F^{-}_{i+1}\right)\cdot n-\frac{1}{2}|\lambda|_{i+1/2}\left(Q^{+}_{i}-Q^{-}_{i+1}\right)\cdot n\,,
\end{equation}
where $|\lambda|_{i+1/2}$ is the maximum eigenvalue of the particular system based on the averages, $Q_{0}$, of the conserved 
variables at the centers of cell $i$ and $i+1$.  The local Lax flux is a well known flux function that can be used in
the discontinuous Galerkin method \cite{Cockburn89}.  For the fluid systems 
$|\lambda|_{i+1/2}=\left(|u_{\alpha}|+\left(\gamma_{\alpha}\,\frac{p_{\alpha}}{\rho_{\alpha}}\right)^{\frac{1}{2}}\right)_{i+1/2}$ 
is used.
For Maxwell's equations $|\lambda|=c$.  The superscripts $+$ and $-$ mean that the $Q$ is evaluated at the 
upper or lower edge of the cell.

\subsection{Limiting}\label{S:Limiting}
High resolution schemes typically use limiting to prevent spurious oscillations near discontinuities and for stabilization of
non-linear systems \cite{Leveque2002}.  Limiters can also be used in the discontinuous Galerkin method, though instead of being
TVD, minmod limiters produce a scheme that is TVDM or TVD in the mean.  This means that the solution is TVD in $Q_{0}$, but not
necessarily in $Q$.

Following the procedure described in \cite{Cockburn89b} the conserved variables $Q$ can be limited in terms of characteristics or in terms of components.  
To limit $Q$ in terms of characteristics the $Q$ are 
first transformed to characteristic variables $g$ where $g=L\,Q$ and $L$ is the left eigenvector
matrix of the flux Jacobian calculated from $Q_{0}$.  The left eigenvector matrix is also applied to the 
differences $L\left(Q^{i+1}_{0}-Q^{i}_{0}\right)=\Delta^{+}g_{0}$ and
$L\left(Q^{i}_{0}-Q^{i-1}_{0}\right)=\Delta^{-}\,g_{0}$.  Limiting is performed directly on transformed variables and then the solution is immediately
transformed back to determine the limited form of $Q_{x}$, 
\begin{equation}\label{E:Limiter}
	\bar{Q}_{x} = L^{-1}\,m\left(g_{x},\Delta^{+}g_{0},\Delta^{-}g_{0}\right)
\end{equation}
where $m$ is the minmod limiter defined by
\begin{equation}\label{E:MinmodLimiter}
	m\left(a,b,c\right)=
	\begin{cases}
	\text{max}\left(a,b,c\right) &\text{if $\text{sign}(a)=\text{sign}(b)=\text{sign}(c)=-$}\\
	\text{min}\left(a,b,c\right) &\text{if $\text{sign}(a)=\text{sign}(b)=\text{sign}(c)=+$}\\
	0 & \text{otherwise}
	\end{cases}\,.
\end{equation}
The minmod limiter Eqn.\thinspace\eqref{E:MinmodLimiter} is typically used for each of the fluid equations Eqn.\eqref{E:ElectronBalanceLaw}\eqref{E:IonBalanceLaw} while the modified minmod limiter can be used to reduce the dissipation
\begin{equation}\label{E:ModifiedMinmodLimiter}
	\bar{m}\left(a,b,c\right)=
	\begin{cases}
	a & \text{if $|a|<M\,dx^{2}$}\\
	m\left(a,b,c\right) & \text{otherwise}\\
	\end{cases}\,,
\end{equation}  
where M is a constant.  Component limiting is done in a similar manner except no transformation is necessary, so that the limiter is directly applied
to the variables $Q$.  Component limiting has the advantage that it is faster than characteristic limiting and it does not introduce machine 
precision errors that can result during the transformation $Q=L^{-1}\left(L\,Q\right)$.  The disadvantage of component limiting is that the approach is not TVDM\cite{Cockburn89b} and unphysical oscillations can appear in the solution.  In this paper characteristic limiting is used.

When a 3rd order DG method is used, two types of limiters can be used.  The first method follows the procedure of second order method and is described 
in \cite{Cockburn90} where if $\bar{Q}_{x}\ne Q_{x}$ then all higher order coefficients are set to zero.  This method is simple to implement and
is used in this paper.

A different and potentially better 3rd order limiter is that of \cite{Biswas94}.  
In this method, the linear terms, $Q_{x}$, $Q_{y}$, $Q_{z}$ are limited in the same way as the second order method while
the higher order terms $Q_{x\,y}$, $Q_{x\,x}$ and $Q_{y\,y}$ are limited as follows.
\begin{equation}
	\bar{Q}_{x\,x}^{i}=m\left[Q_{x\,x}^{i},\frac{1}{2}\left(Q_{x}^{i+1}-Q_{x}^{i}\right),\frac{1}{2}\left(Q_{x}^{i}-Q_{x}^{i-1}\right)\right]
\end{equation}
and
\begin{equation}
	\bar{Q}_{y\,y}^{j}=m\left[Q_{y\,y}^{j},\frac{1}{2}\left(Q_{y}^{j+1}-Q_{y}^{j}\right),\frac{1}{2}\left(Q_{y}^{j}-Q_{y}^{j-1}\right)\right]
\end{equation}
finally, the term $Q_{x\,y}$ is limited by setting it to zero if either $Q_{x\,x}$ or $Q_{y\,y}$ is limited.  In \cite{Cockburn90} it was
suggested that the high order terms could be limited by simply setting them to zero if the linear terms are limited.  The justification is
that oscillations in the higher order terms would only be important when oscillations in the linear terms exist.
\subsection{Stability}
The stability limits of the numerical algorithm just described are defined by the highest oscillation frequency
of the system or by the CFL condition based on the speed of light.  Typically the highest oscillation frequency
is the electron plasma frequency $w_{p\,e}=\left(\frac{n_{e} q_{e}^{2}}{\epsilon_{0}\,m_{e}}\right)^{\frac{1}{2}}$
and a time step is chosen for which the time integration scheme is stable to this frequency of 
oscillation, this time step is typically $\Delta t<\frac{1}{w_{p\,e}}$.  When the CFL condition dominates the time step
 $\Delta t<\frac{1}{6}\frac{\Delta x}{c}$
is used for the second order spatial discretization in 2D and $\Delta t<\frac{1}{10}\frac{c}{\Delta x}$ for the third order
spatial discretization in 2D.  
\section{Simulations}\label{S:Simulations}
The two-fluid system describes many dispersive waves including the electron acoustic
wave.  In a plasma the electron acoustic wave is coupled to the plasma frequency producing
a wave that is essentially stationary for sufficiently long wavelengths or sufficiently
cold plasmas.  In the following a dispersion relation for the electron acoustic wave in
a warm plasma is derived from linearized two-fluid equations.  The dispersion relation is used
to calculate an analytic solution to the propagation of an approximate square pulse in a two-fluid
plasma.  The numerical two-fluid solution using the 2nd and 3rd order discontinuous Galerkin method
is compared to the analytic solution using various grid resolution.  The order accuracy of the
algorithm in the $L_{2}$ norm is calculated from these results.

In real kinetic plasmas damping such as Landau damping may result in 
the damping of waves observed in simulations such as this.  These results are meant to 
illustrate verification that we are solving the correct system in addition to showing 
we can achieve the desired accuracy.

\begin{figure}
	\begin{center}
	\scalebox{0.6}{\includegraphics{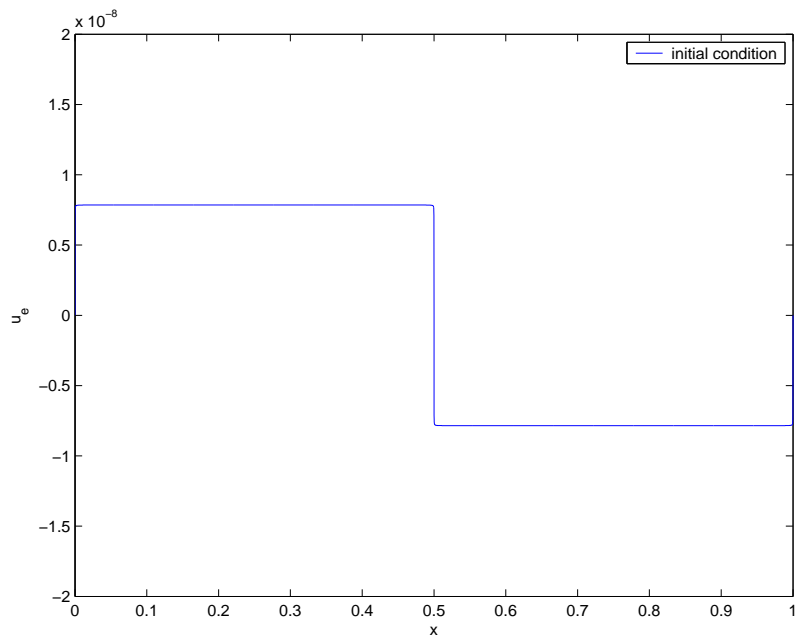}}
	\end{center}
	\caption{Electron velocity of a square electron acoustic square pulse at time $t=0$}
	\label{F:Example-initial-condition}
\end{figure}

\begin{figure}
	\begin{center}
	\scalebox{0.6}{\includegraphics{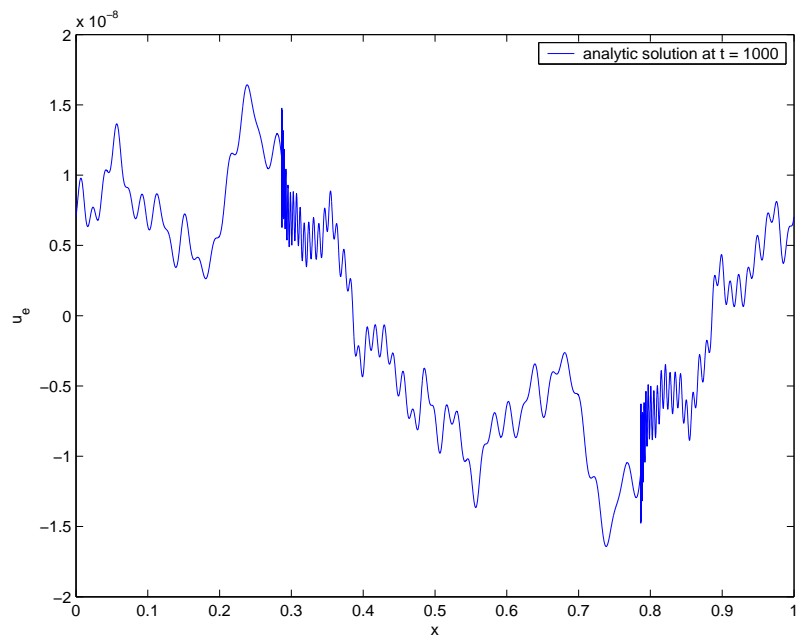}}
	\end{center}
	\caption{Analytic solution of the electron velocity of a square electron acoustic pulse at time $t=1000$.  This plot
	illustrates the dispersive, non-diffusive nature of the two-fluid system that makes it numerically difficult
	even in the linear regime.
	The high frequency modes still contribute significantly to the amplitude of the solution late in
	time and they always will because there is no physical diffusion. The dispersive, non-diffusive nature
	of the two-fluid system can make it appear that our numerical solutions suffer from 
	significant numerical dispersion errors when they are actually correctly capturing 
	dispersion described by the model.}
	\label{F:Example-final-condition}
\end{figure}

\begin{figure}
	\begin{center}
	\scalebox{0.6}{\includegraphics{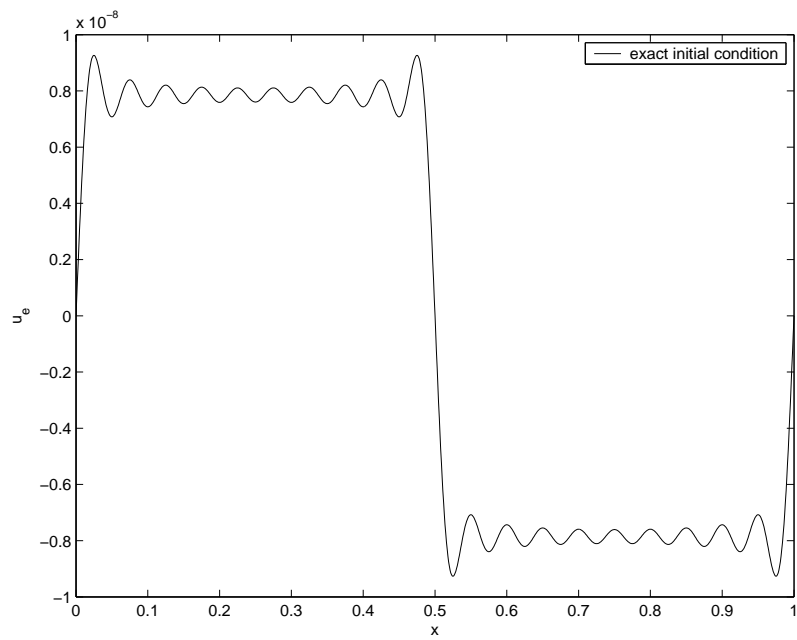}}
	\end{center}
	\caption{Exact electron x velocity at t=0.  Initial conditions are chosen
	so that all waves travel to the left as time increases.  Since the waves are
	dispersive the initial ``square wave'' disappears when the solution is allowed
	to evolve.}\label{F:EA-initial-condition}
\end{figure}

Assume infinitely massive ions with a background number density $n_{0}$ for both
electrons and ions and charge $q_{i}=-q_{e}$.
Furthermore, assume background electron and ion pressures $P_{0}$ while all other background quantities
are zero.  A perturbed electron velocity $u_{e}^{1}=U_{0}\,e^{i\left(k_{n}\,x+w_{n}\,t\right)}$ is assumed.  
Corresponding perturbed electric field, density and pressure profiles can be derived from Poisson's equation,
the continuity equation and the energy equation so that the perturbed electric field 
$E_{x\,n}^{1}=\frac{i}{\epsilon_{0}\,w_{n}}n_{0}q_{e}\,u_{e}^{1}$, perturbed electron pressure,
$P_{e\,n}^{1}=-\left(\frac{k_{n}}{w_{n}}\right)\gamma_{e}\,P_{0}\,u_{e}^{1}$,
and perturbed electron density, $\rho_{e\,n}^{1}=-\left(\frac{k_{n}}{w_{n}}\right)\rho_{0\,e}\,u_{e}^{1}$.
The electron acoustic dispersion relation is
\begin{equation}
	w_{n}=\pm\left[\left(\frac{\gamma_{e}\,P_{0}}{\rho_{0\,e}}\right)\,k_{n}^{2}+
	\left(\frac{n_{0}\,q_{e}^{2}}{\epsilon_{0}\,m_{e}}\right)\right]^{\frac{1}{2}}\,.
\end{equation}
The positive root defines waves that travel to the left.  A square pulse on
a periodic domain is defined by taking linear combinations of these waves,
\begin{equation}
	u_{e}^{1}\left(x,t\right)= -U_{0}\sum_{n=0}^{\infty}\,\frac{i}{2n+1}e^{i\left(\,k_{n}\,x+w_{n}\,t\right)}\,.
\end{equation}
As a practical matter, an approximate square pulse is used since the
high wave numbers cannot be resolved numerically unless the spatial resolution is sufficiently
high.  Figures \ref{F:Example-initial-condition} and \ref{F:Example-final-condition} illustrate
this dramatically.  Figure \ref{F:Example-initial-condition} shows the initial square pulse initialized
with 5000 wave modes.  Figure \ref{F:Example-initial-condition} shows the analytic solution after $t=1000$
steps.  Since there is no dissipation in the system and the system is dispersive the high frequency modes
in the initial conditions play an important role in the final solution; this makes the issue of convergence
in shock-type problems that start out with discontinuities difficult to assess.  As a result, in the simulations and 
analytic solutions that follow, $n=9$ is the highest mode included
in the expansion and $k_{n}=2\,\pi\,n$.  Finally, only the real part of all perturbed quantities are
used in the initial conditions, thus,
\begin{equation}
	E_{x}^{1}\left(x,t\right)=\sum_{n=0}^{9}E_{x\,n}^{1}
\end{equation}
\begin{equation}
	P_{e}^{1}\left(x,t\right)=\sum_{n=0}^{9}P_{e\,n}^{1}
\end{equation}
\begin{equation}
	\rho_{e}^{1}\left(x,t\right)=\sum_{n=0}^{9}\rho_{e\,n}^{1}
\end{equation}
In this simulation $q_{i}=-q_{e}=10$, $\epsilon_{0}=1$, $n_{0}=1$, $P_{0}=1$, $m_{e}=1$, 
$m_{i}=\infty$, $\gamma_{e}=2$ for convenience.  To ensure that the solution is in the
linear regime $U_{0}$ must be set to a small value.  For these simulations $U_{0}=1\times 10^{-8}$.  The domain is 
periodic with length $1$.  Electromagnetic waves do not exist in this problem, but the speed
of light $c=1$.  The simulations are run to time $t=3$ at several different resolutions
and 20,000 time steps are taken for the highest resolution simulation which has 320 cells.  
The initial conditions for the electron x-velocity 
perturbation are shown in figure \ref{F:EA-initial-condition}.  An approximate square wave is used to 
excite several wave modes to test the algorithms performance effectively.  A problem in the
linear regime is used for two reasons.  First of all, analytic solutions exist and are easy
to calculate, secondly, in the linear regime the limiters can be turned off so the solution
can be observed without the added dissipation which can reduce overall accuracy making it
much easier to compute numerically the accuracy of the algorithm.  Ultimately, problems in
the non-linear regime, problems with large $U_{0}$ for example, require limiters and the overall
accuracy of the numerical solution is reduced.  The design of effective limiters may be the
most important problem in gaining computational efficiency from 3rd order or higher discontinuous
Galerkin methods for the two-fluid system when compared to the 2nd order method.
\clearpage

\begin{figure}
	\begin{center}
	\scalebox{0.6}{\includegraphics{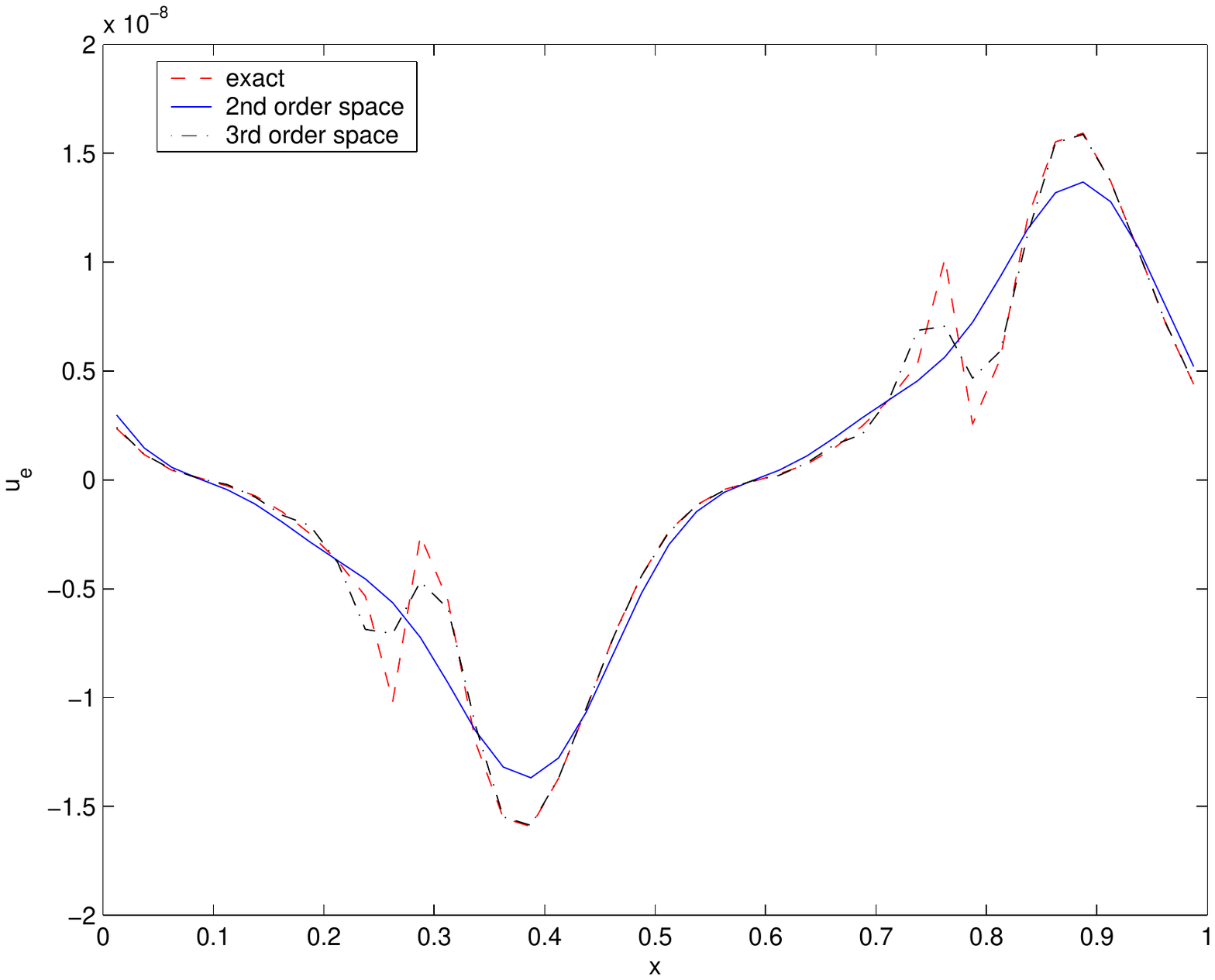}}
	\end{center}
	\caption{Numerical solution using 2nd and 3rd order discontinuous Galerkin spatial
	discretizations with a 3rd order time discretization compared to the exact solution
	at t=3.  The grid has $40$ cells so the highest wave number mode is barely resolved
	with the 2nd order method.  The 2nd and 3rd order solutions differ substantially
	from the analytic solution.}\label{F:EA-converge-40}
\end{figure}

\begin{figure}
	\begin{center}
	\scalebox{0.6}{\includegraphics{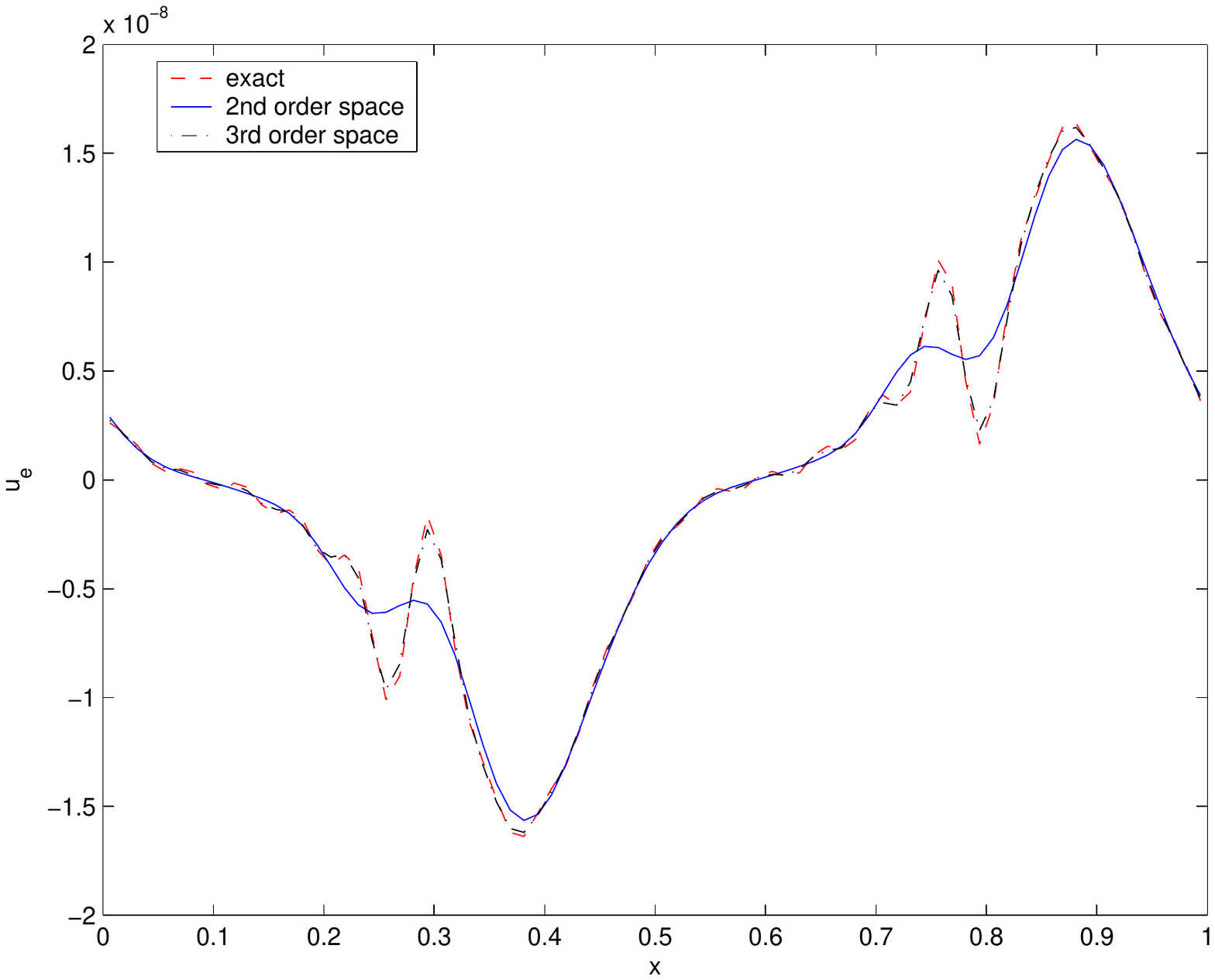}}
	\end{center}
	\caption{Numerical solution using 2nd and 3rd order discontinuous Galerkin spatial
	discretizations with a 3rd order time discretization compared to the exact solution
	at t=3.  The grid has $80$ cells across the domain.  The 2nd order solution differs
	substantially from the analytic solution, only capturing the lower order modes.  The 
	3rd order solution matches the analytic solution well.}\label{F:EA-converge-80}
\end{figure}

\begin{figure}
	\begin{center}
	\scalebox{0.6}{\includegraphics{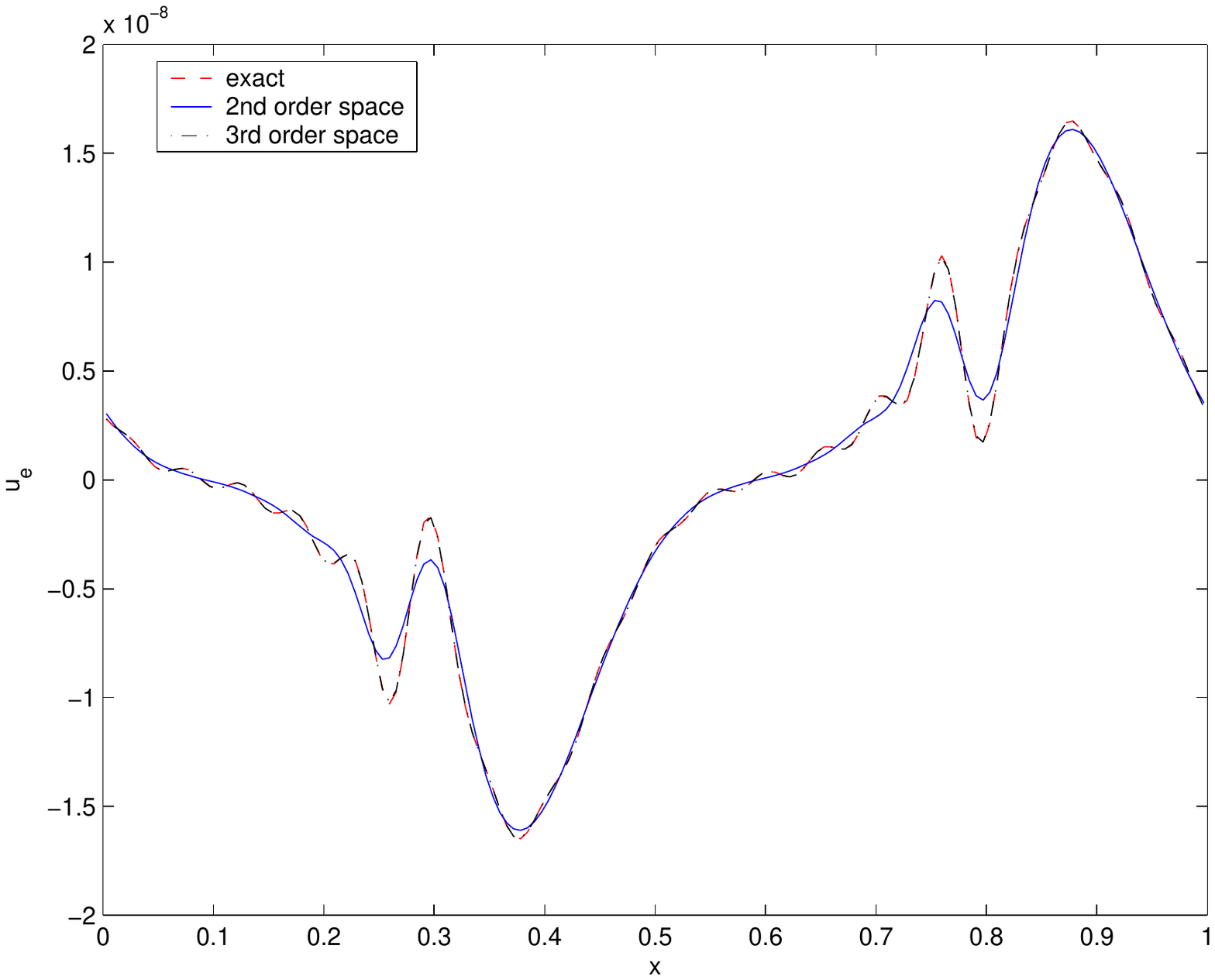}}
	\end{center}
	\caption{Numerical solution using 2nd and 3rd order discontinuous Galerkin spatial
	discretizations with a 3rd order time discretization compared to the exact solution
	at t=3.  The grid has $160$ cells across the domain.  The 2nd order solution resolves
	the high order modes at this resolution.  At higher resolution the numerical solutions
	are visually indistinguishable from the analytic solution.}\label{F:EA-converge-160}
\end{figure}

\begin{figure}
	\begin{center}
	\scalebox{0.6}{\includegraphics{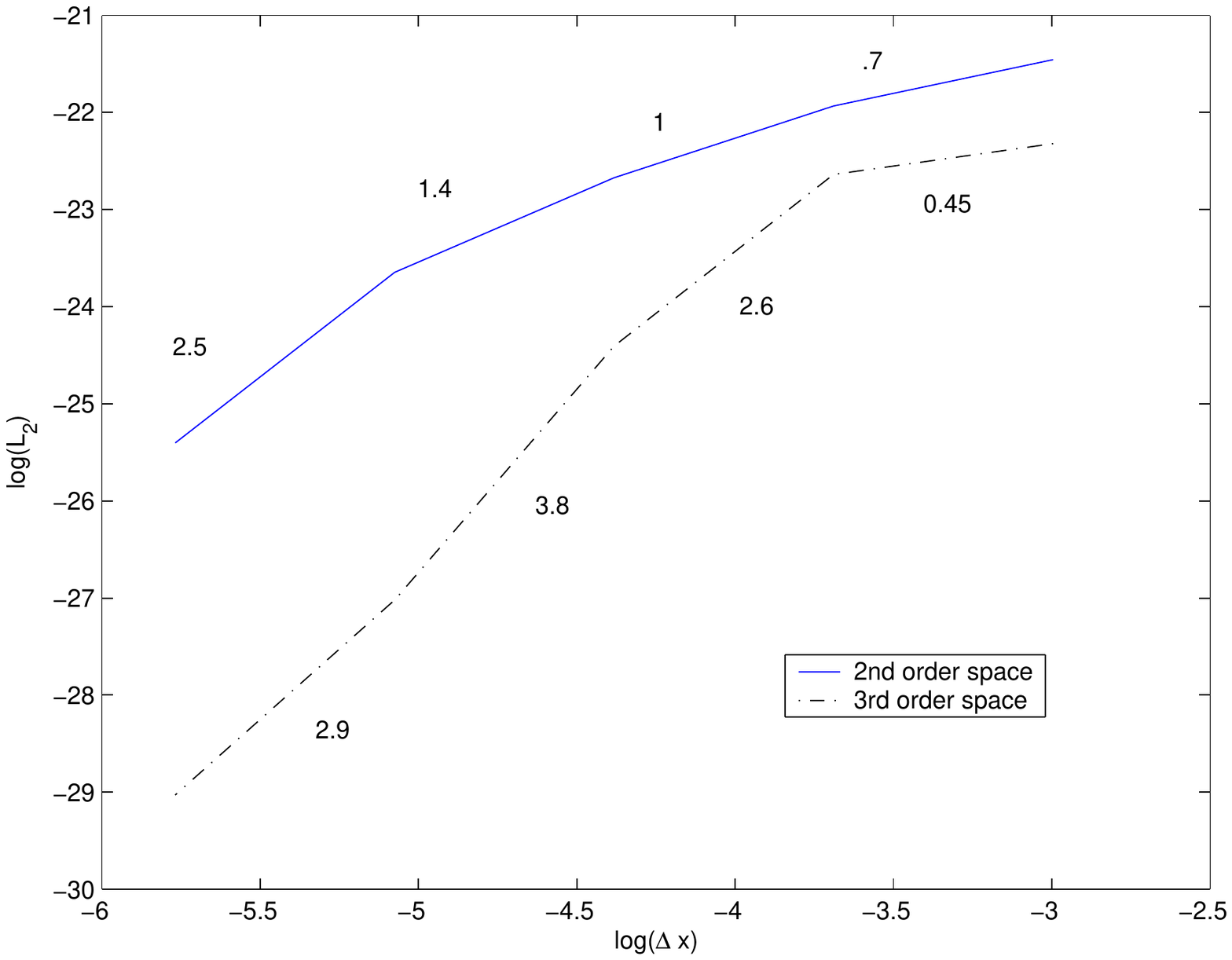}}
	\end{center}
	\caption{Natural log of the $L_{2}$ norm verses natural log of grid spacing for
	the numerical solutions to the electron acoustic wave dispersion problem.  The solutions
	were calculated using 2nd and 3rd order discontinuous Galerkin spatial discretizations with
	a 3rd order Runge-Kutta time discretization.  The numbers near each line give the slope of the
	line and hence the measured order of accuracy of the scheme.  The 3rd order method gives
	substantially better accuracy than the second order method.  Grid resolutions used to construct
	this plot are $1/20$, $1/40$, $1/80$, $1/160$, and $1/320$.  The measured accuracy was
	computed at $t=3$.}\label{F:EA-log-error}
\end{figure}

After $3$ time units the square pulse shape has disappeared due to wave dispersion.  Plots
of the numerical solution verses the analytic solution at various grid resolutions are
shown in figures \ref{F:EA-converge-40}, \ref{F:EA-converge-80}, and \ref{F:EA-converge-160} at $3$
time units.  In figure \ref{F:EA-converge-40} there are $40$ cells in the domain and the 3rd order
method shows evidence of resolving the highest order mode.  The 2nd order method only captures
the bulk features.
In figure \ref{F:EA-converge-80} there are $80$ cells in the domain and the 3rd order method 
captures the amplitude of the highest modes
and matches the analytic solution very well.  The 2nd order method still struggles to resolve
the highest mode (note the solution at the two skinniest spikes).
In figure \ref{F:EA-converge-160} there are $160$ cells in the domain and the 2nd order method 
still does not match the amplitude of the highest order mode and does not match the amplitudes 
much better than the 3rd order method at $40$ grid cells.
Figure \ref{F:EA-log-error} shows a plot of the convergence history of the numerical solutions
along with calculated order of accuracy.  In both cases the calculated order of accuracy varies 
from less than $1$ for low resolution where the high frequency modes are barely resolved to better
than the order of the scheme when the solution is nearly converged.

The 3rd order method performs substantially better than the 2nd order in this linearized problem when
limiter are not needed.  In particular, the 3rd order method preserves amplitude much better than the 2nd
order, this same phenomena has been observed in certain equilibrium type non-linear problems.
\clearpage
\subsection{Two-Fluid Electromagnetic Plasma Shock}
The two-fluid electromagnetic plasma shock is an extension of the Brio and Wu shock \cite{Brio88} to the two-fluid plasma model.  The simulation
was first performed in \cite{Shumlak2003,Loverich2003} and is used in the current paper as a benchmark.  The ideal two-fluid system has no
dissipative terms however an artificial viscosity exists due to the numerical discretization.  Wave steepening of the two-fluid solution is limited
by physical wave dispersion, when wave dispersion is not sufficient to limit the steepening, artificial viscosity limits the steepening.  In real collisionless
shocks in experiments and in space, kinetic effects limit the wave steepening when dispersion is not sufficient to limit the steepening.  This simulation is
meaningful since it illustrates the range of physics that the two-fluid system describes.

\begin{figure}[!ht]
	\begin{center}
	\scalebox{0.6}{\includegraphics{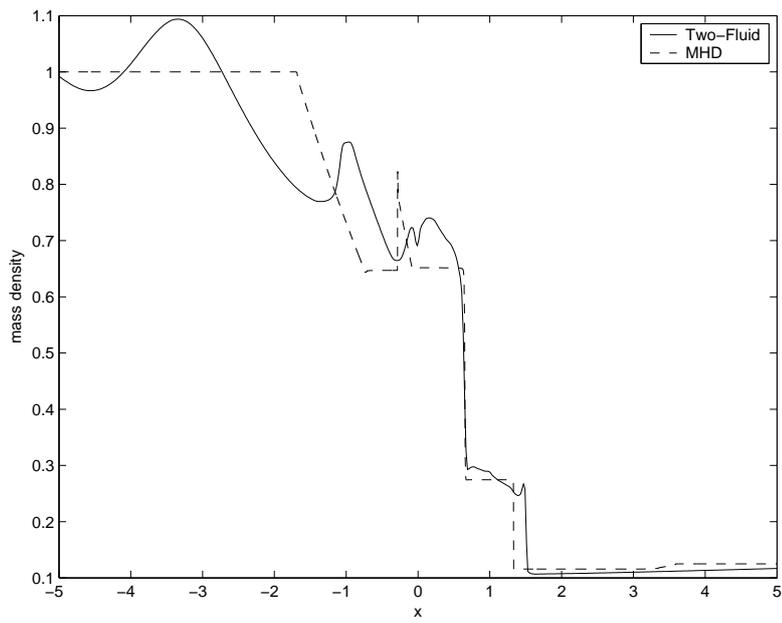}}
	\end{center}
	\caption{Electromagnetic shock solution using the two-fluid equations and the MHD equations at $t=0.01\tau_{c}$.  At this
	time the domain spans $10$ ion Larmor radii or $1000$ Debye lengths.  It is in this regime that the two-fluid solution differs most
	significantly from the MHD or the ``gas dynamic" solution.  This regime has practical applications to Z-Pinches and
	FRC's due to the weak magnetization of the ions.}\label{F:MHDq10}
\end{figure}

\begin{figure}
	\begin{center}
	\scalebox{0.6}{\includegraphics{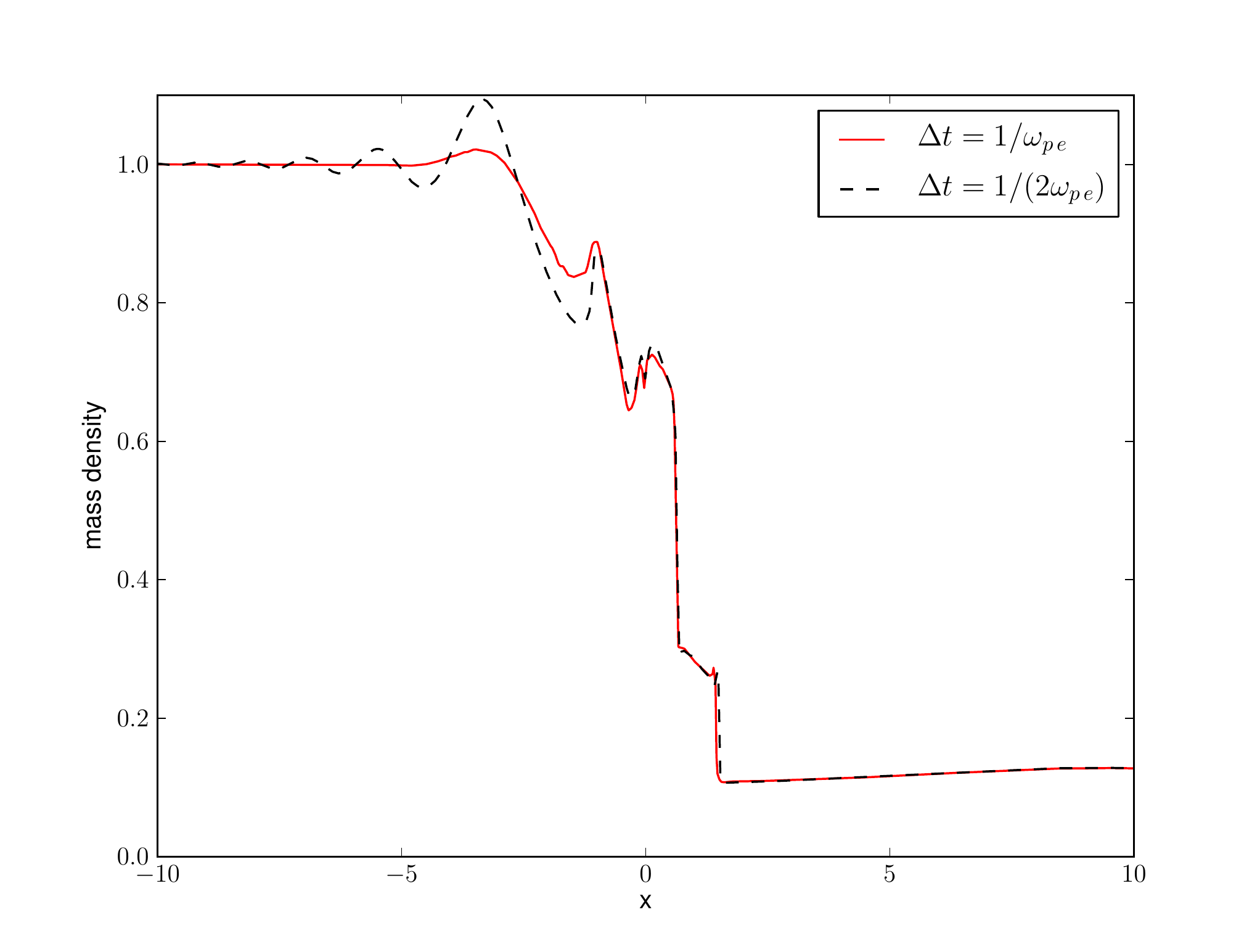}}
	\end{center}
	\caption{Comparison of $10$ ion Larmor radius solution when the plasma frequency is just resolved compared to a solution with half the time step.  Time step close to the stability limit
	significantly damps the high frequency waves virtually eliminating the trailing wave train.}\label{F:MHDq10compare}
\end{figure}

\begin{figure}
	\begin{center}
	\scalebox{0.6}{\includegraphics{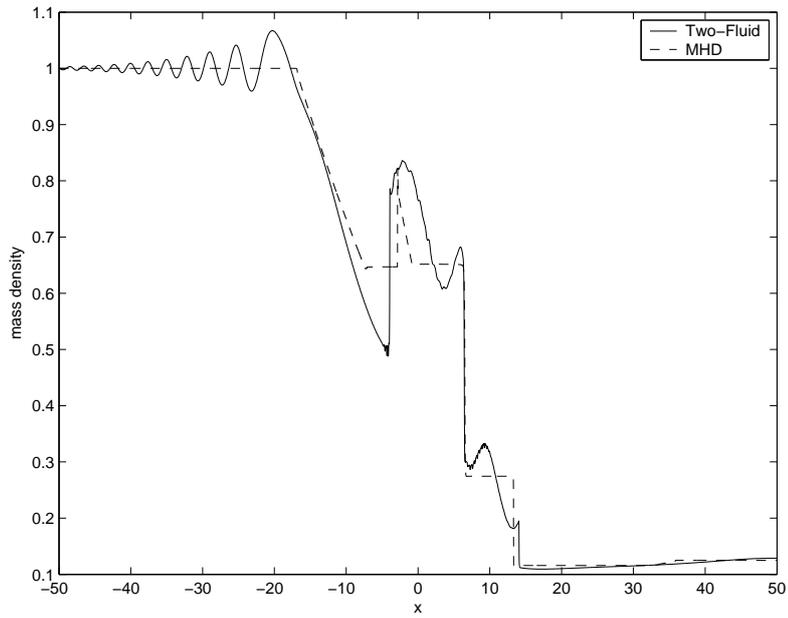}}
	\end{center}
	\caption{Electromagnetic shock solution using the two-fluid equations and the MHD equations at $t=0.1\tau_{c}$.  At this
	time the domain spans $100$ ion Larmor radii or $10000$ Debye lengths.  Major differences from previously published results (Figure 9 in \cite{Shumlak2003}
	include the large oscillations to the left of the rarefaction wave which are dispersive plasma waves.  The differences are a result of the higher grid resolution and
	better accuracy of the algorithm used in this paper.}\label{F:MHDq100}
\end{figure}

\begin{figure}
	\begin{center}
	\scalebox{0.6}{\includegraphics{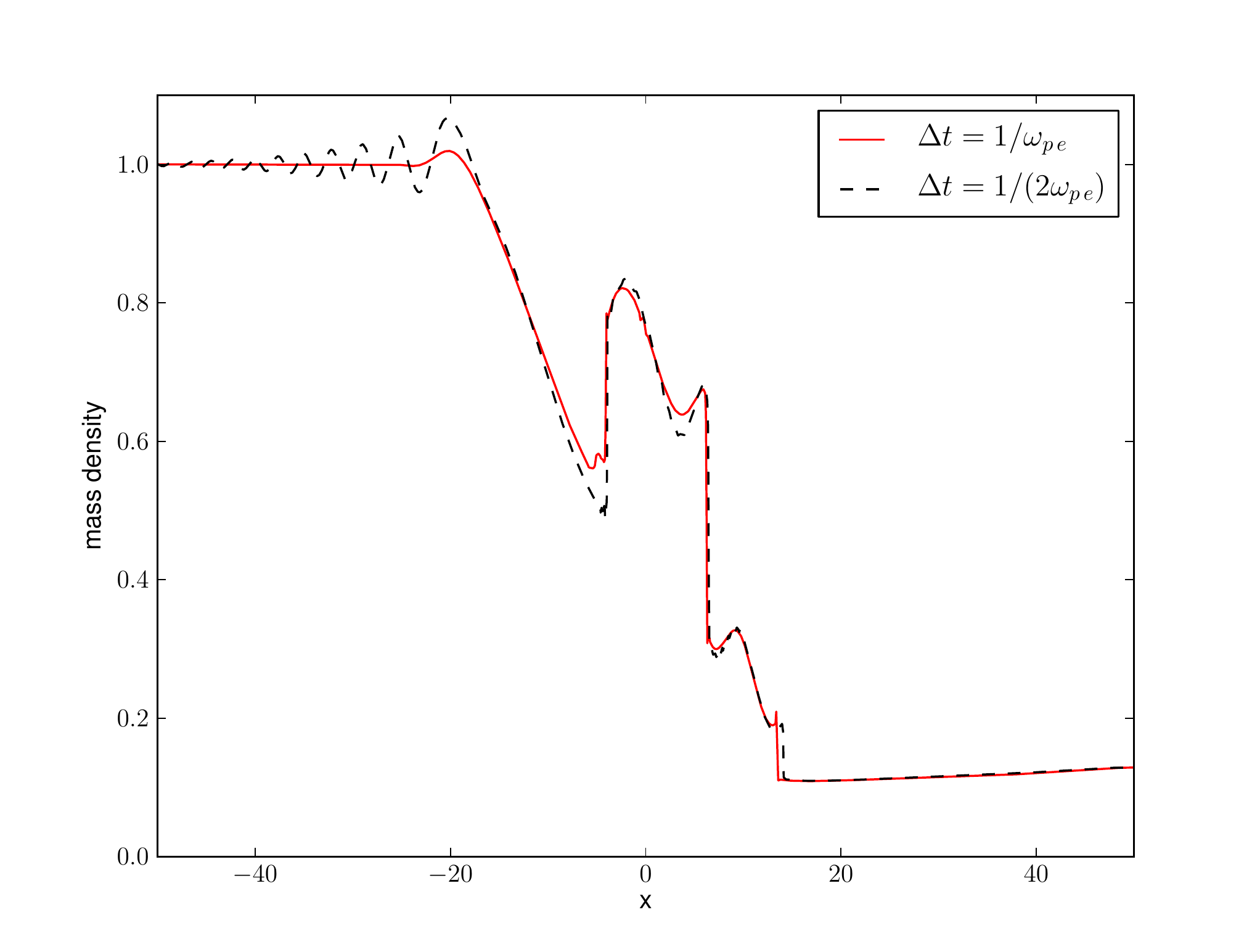}}
	\end{center}
	\caption{Comparison of $100$ ion Larmor radius solution when the plasma frequency is just resolved compared to a solution with half the time step.}\label{F:MHDq100compare}
\end{figure}

\begin{figure}
	\begin{center}
	\scalebox{0.6}{\includegraphics{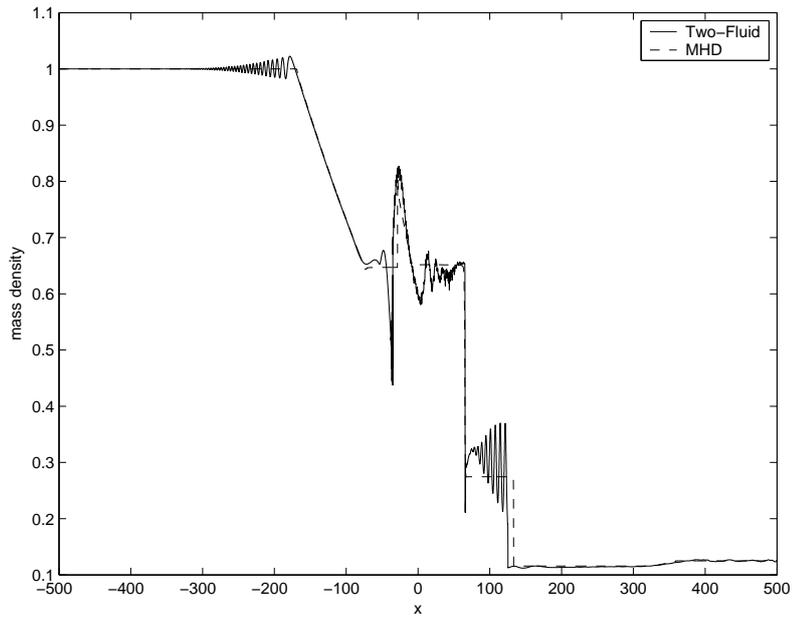}}
	\end{center}
	\caption{Electromagnetic shock solution using the two-fluid equations and the MHD equations at $t=1\tau_{c}$. At this point in the
	simulation the solution is very MHD like.  However, key differences remain as a result of the fact that the two-fluid system models dispersive MHD waves.  The 
	most major differences are the post shock and post rarefaction wave oscillations.
	Both look numerical, but these waves are resolved in several hundred grid cells and are a result of dispersive plasma waves described in the model.  Moving from lower resolution to higher resolution runs
	show more high frequency waves with less diffusion.  This simulations was the highest resolution performed using 50,000 cells.  In this regime the Hall MHD model should be used since if one uses the full two-fluid model one needs to approximately resolve the Debye length scales to get good shock solutions.  The desire to resolve the Debye length in this simulation means that extremely high resolution was used.}\label{F:MHDq1000}
\end{figure}

\begin{figure}
	\begin{center}
	\scalebox{0.6}{\includegraphics{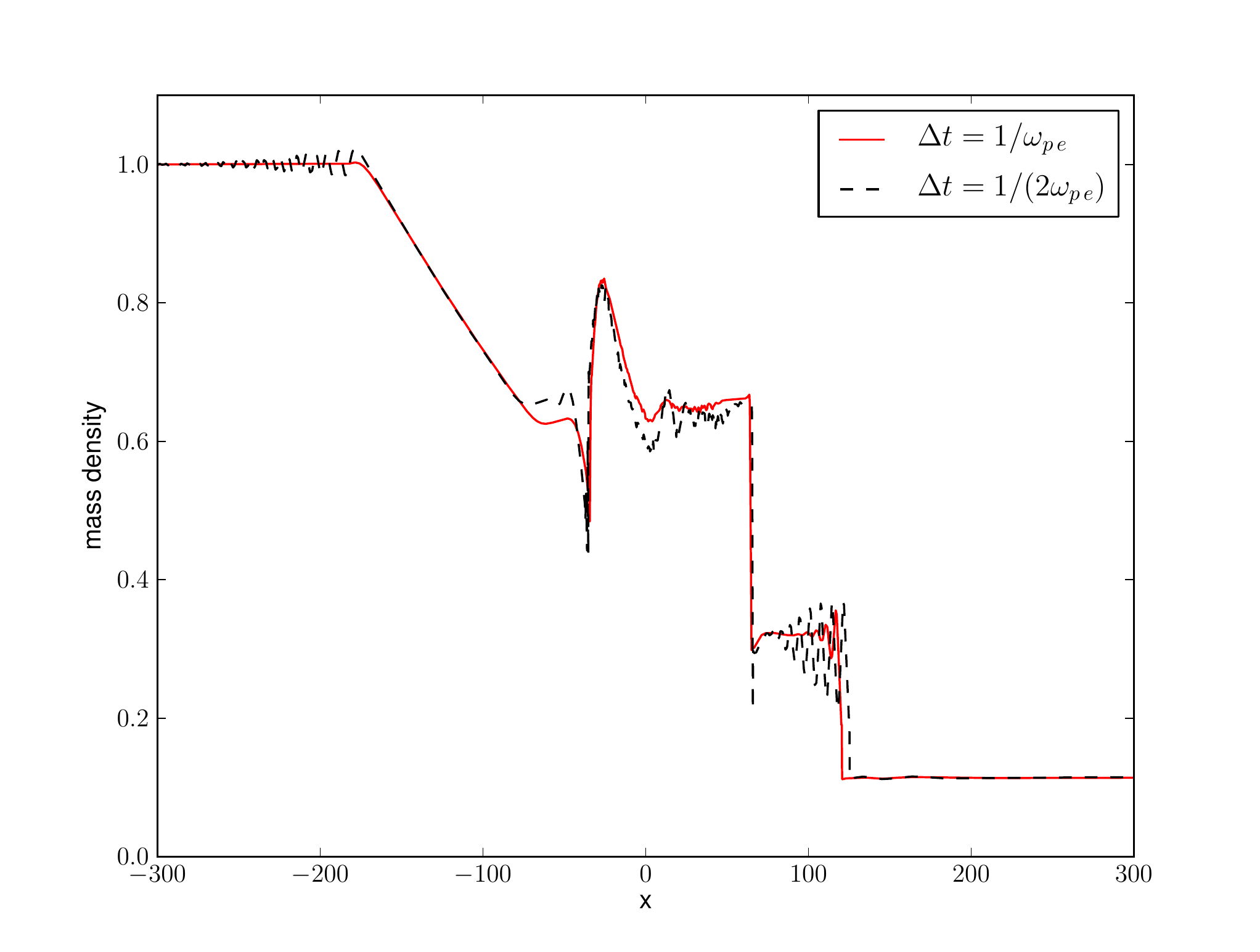}}
	\end{center}
	\caption{Comparison of $1000$ ion Larmor radius solution when the plasma frequency is just resolved compared to a solution with half the time step.  It's very important to resolve he plasma frequency in order to get the full dynamics of this simulation.}\label{F:MHDq1000compare}
\end{figure}

In this paper the shock will be presented differently than in \cite{Shumlak2003}.
The initial discontinuity is allowed to evolve in time until the shock structure spans $1000 r_{g\,i}$ 
where $r_{g\,i}$ is the ion Larmor radius.  Time is measured
in terms of light transit times across the entire domain, $\tau_{c}=\frac{1000 r_{g\,i}}{c}$.  Snapshots
of the shock earlier in time correspond to larger characteristic ion Larmor radius, $\frac{r_{g\,i}}{L}<1000$ 
where $L$ is the span of the shock, 
and so the solution evolves from a ``gas dynamic'' 
regime of short time scales and large characteristic ion Larmor radius $\frac{r_{g\,i}}{L}\gg 1$ to an ``MHD'' 
regime of long 
time scales and small characteristic Larmor radii $\frac{r_{g\,i}}{L}\ll 1$.
Parameters used are, $q_{i}=10$, $q_{e}=-10$, and $\epsilon_{0}=1$, 
$\mu_{0}=1$, $c=1$, $\gamma_{e}=\gamma_{i}=\frac{5}{3}$,
$m_{i}=1$, $m_{e}=\frac{1}{1836}$.
The initial conditions on the left half of the domain are given by,
\begin{equation}
	\left(
	\begin{array}{c}
	n_{e}\\
	U_{e\,x}\\
	U_{e\,y}\\
	U_{e\,z}\\
	P_{e}\\
	n_{i}\\
	U_{i\,x}\\
	U_{i\,y}\\
	U_{i\,z}\\
	P_{i}\\
	B_{x}\\
	B_{y}\\
	B_{z}\\
	E_{x}\\
	E_{y}\\
	E_{z}\\
	\end{array}
	\right)_{left}=
	\left(
	\begin{array}{c}
	1.0\\
	0\\
	0\\
	0\\
	0.5\times 10^{-4}\\
	1.0\\
	0\\
	0\\
	0\\
	0.5\times 10^{-4}\\
	0.75\times 10^{-2}\\
	1.0\times 10^{-2}\\
	0\\
	0\\
	0\\
	0\\
	\end{array}
	\right),\quad \text{and}\quad
	\left(
	\begin{array}{c}
	n_{e}\\
	U_{e\,x}\\
	U_{e\,y}\\
	U_{e\,z}\\
	P_{e}\\
	n_{i}\\
	U_{i\,x}\\
	U_{i\,y}\\
	U_{i\,z}\\
	P_{i}\\
	B_{x}\\
	B_{y}\\
	B_{z}\\
	E_{x}\\
	E_{y}\\
	E_{z}\\
	\end{array}
	\right)_{right}=
	\left(
	\begin{array}{c}
	0.125\\
	0\\
	0\\
	0\\
	0.05\times 10^{-4}\\
	0.125\\
	0\\
	0\\
	0\\
	0.05\times 10^{-4}\\
	0.75\times 10^{-2}\\
	1.0\times 10^{-2}\\
	0\\
	0\\
	0\\
	0\\
	\end{array}
	\right)\,.
\end{equation}
The spatial units of figures \ref{F:MHDq10}, \ref{F:MHDq10compare}, \ref{F:MHDq100}, \ref{F:MHDq100compare}, \ref{F:MHDq1000}, \ref{F:MHDq1000compare} are measured in 
ion Larmor radii $r_{g\,i}$ based on the initial conditions
in the left half of the domain.  The Debye length is $\lambda_{d}=\frac{1}{100}\,r_{g\,i}$ also based on the initial conditions on the left half of the domain.  The results 
in figure \ref{F:MHDq10} correspond to a less accurate solution to that published in \cite{Shumlak2003} figure 8.  The domain of figure \ref{F:MHDq10} has 
only 500 cells in the domain corresponding to 1000 degrees of freedom (5,000 time steps to reach this point in the simulation) while the published solution using a finite volume method has 4000 cells or 4000 degrees of freedom.  In figure \ref{F:MHDq100} the solution is higher resolution than the corresponding solution published in figure 9 of  \cite{Shumlak2003} as can be seen by the resolution of the oscillations to the left 
of the rarefaction wave.  At this scale there are 5000 cells in the domain corresponding to 10,000 degrees of freedom (50,000 time steps to reach this point in the simulation) while there are 4000 degrees of freedom in the previously published solution.  
At the final time figure \ref{F:MHDq1000}, the solution has moved beyond those published and 
significant oscillations to the left of the shock are observed.  Most of these waves are resolved in several hundred 
grid cells, and the entire domain of figure \ref{F:MHDq1000} is 50,000 cells (500,000 time steps to reach this point in the simulation).

Due to the dispersive, non-diffusive nature of the system, the shocks solutions show more complexity as resolution is increased in time and space.  To illustrate this  figures \ref{F:MHDq10compare}, \ref{F:MHDq100compare},
\ref{F:MHDq1000compare} show solutions using a time step based $\Delta t=1/\omega_{p\,e}$ compared to the solution with $\Delta t=1/(2\omega_{p\,e})$.  These results show the large amount of damping in the $\Delta t=1/\omega_{p\,e}$ case that results from temporally unresolved simulations, however, many of the essential features of the solution still remain.  Similarly, increasing the spatial resolution adds higher frequency dispersive waves without significantly changing the overall shape.  The final solution \ref{F:MHDq1000} took roughly two days on 64 processors so higher resolution solutions were not attempted.
\clearpage

\subsection{Magnetic Reconnection}
In ideal MHD the fluid is frozen to the magnetic field lines and this prevents one field line from connecting
with another.  The addition of non-ideal terms such as resistivity results in fluid moving across field lines allowing magnetic 
reconnection to occur.  Classical resistivity leads to much slower reconnection rates than are observed in collisionless space and fusion 
plasmas.  Non-ideal, collisionless terms such as electron inertia and the electron pressure gradient are partially responsible 
for the fast reconnection that is observed to occur in the earths magnetotail and fusion plasmas.
In this section we show that the two-fluid algorithm developed in this paper produces magnetic reconnection 
rates that agree with those described of the GEM challenge magnetic reconnection problem as 
described in \cite{Birn2001}.

The GEM challenge magnetic reconnection problem is non-dimensionalized as in
\cite{Birn2001} where lengths are normalized by the ion
inertial length $d=c/w_{p\,i}=c\,\left(\frac{e^{2}\,n_{0}}{\epsilon_{0}\,m_{i}}\right)^{-\frac{1}{2}}$
time is non-dimensionalized by the ion-cyclotron time $\frac{m_{i}}{e\,B_{0}}$ where $B_{0}$ is the
magnetic field at infinity.  The velocities are normalized by the Alfven velocity 
$V_{a}=\left(\frac{B_{0}^{2}}{\mu_{0}\,m_{i}\,n_{0}}\right)^{\frac{1}{2}}$.  Finally current density 
is non-dimensionalized by $J_{0}=\frac{B_{0}\,w_{p\,i}}{\mu_{0}\,c}$ and E by $E_{0}=V_{a}\,B_{0}$.
The domain is $\left(-6.4\,d,6.4\,d\right)$ and the simulation is run out to $40/w_{c\,i}$.  Conducting walls
are used on the $y$ boundaries and periodic boundaries are used on the $x$ boundaries.  $\lambda=0.5\,d$, the ion to electron mass ratio 
is taken to be $25$ and the specific heat ratio $\gamma=\frac{5}{3}$.  The speed of light is $c=10\,V_{a}$.
The reconnection rates do not change noticeably when the ratio of the speed of light to the Alfven speed is 
increased to $c=100\,V_{a}$ as was done in \cite{Loverich2005}.  The initial number densities are given by,
\begin{equation}
	n_{e}=n_{i}=n_{0}\left(\frac{1}{5}+\text{sech}^{2}\left(\frac{y}{\lambda}\right)\right)\,.
\end{equation}
The electron and ion temperatures differ slightly, but are constant throughout the domain, this
gives the following electron pressure, $P_{e}$,
\begin{equation}
	P_{e}=\frac{1}{12\mu_{0}}B_{0}^{2}\,\frac{n_{e}}{n_{0}}
\end{equation}
and ion pressure $P_{i}$
\begin{equation}
	P_{i}=\frac{5}{12\mu_{0}}B_{0}^{2}\,\frac{n_{i}}{n_{0}}.
\end{equation}
The electron and ion pressure balance the magnetic field which is given by	
\begin{equation}
	B_{x}=B_{0}\,\text{tanh}\left(\frac{y}{\lambda}\right)+\frac{B_{0}}{10}\frac{\pi}{L_{x}}\cos\left(\frac{2\pi\,x}{L_{x}}\right)\sin\left(\frac{\pi\,y}{L_{y}}\right)
\end{equation}
\begin{equation}
	B_{y}=\frac{B_{0}}{10}\left(\frac{2\pi}{L_{x}}\right)\sin\left(\frac{2\pi\,x}{L_{x}}\right)\cos\left(\frac{\pi\,y}{L_{y}}\right)
\end{equation}
The magnetic field is in equilibrium with with the electron current $J_{z\,e}$,
\begin{equation}
	J_{z\,e}=\frac{\mu_{0}\,B_{0}}{\lambda}\,\text{sech}^{2}\left(\frac{x}{\lambda}\right)\,.
\end{equation}
\begin{figure}
	\begin{center}
	\scalebox{0.7}{\includegraphics{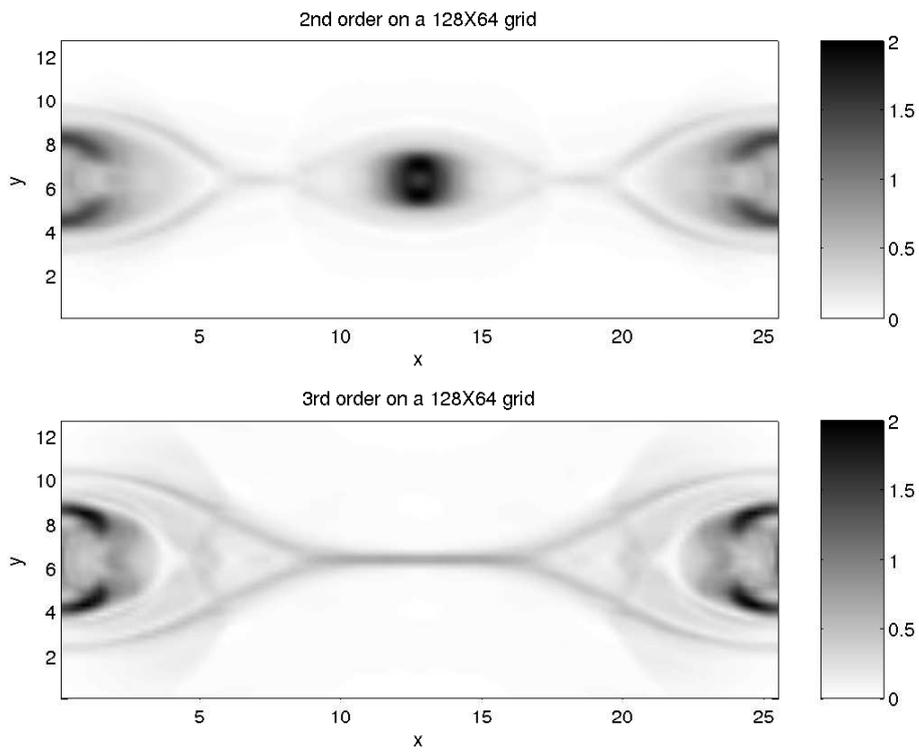}}
	\end{center}
	\caption{GEM challenge comparing $|J_{z}|$ using the 2nd and 3rd order method on a $128 \times 64$ grid at time $t=25/w_{c\,i}$.  At this resolution
	an island forms in the 2nd order method and grows as the simulation progresses.  Both methods use 3rd order TVD Runge-Kutta time stepping.}\label{F:LowResReconnect}
\end{figure}
\begin{figure}
	\begin{center}
	\scalebox{0.7}{\includegraphics{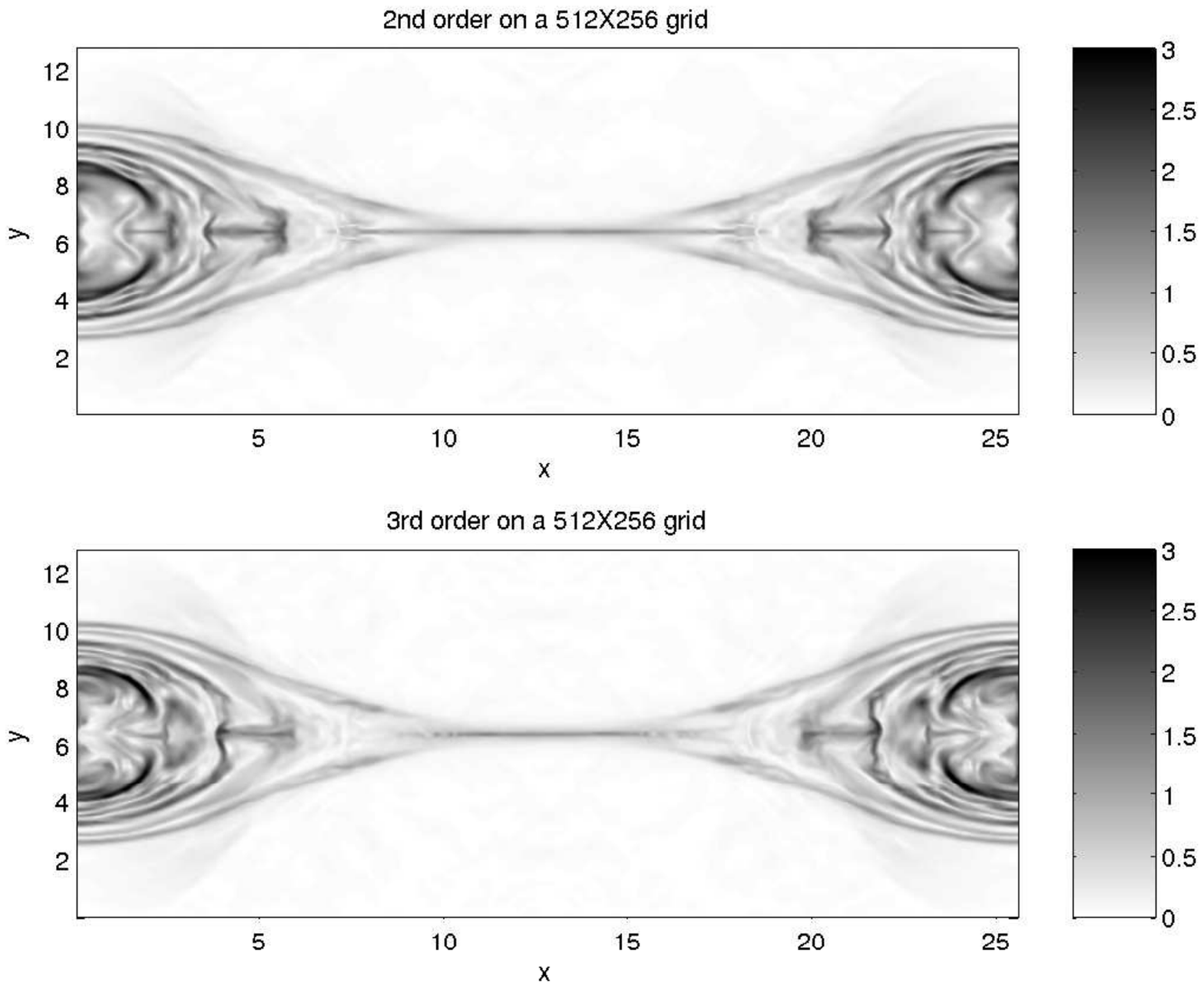}}
	\end{center}
	\caption{GEM challenge comparing $|J_{z}|$ using the 2nd and 3rd order method on a $512 \times 256$ grid at time $t=25/w_{c\,i}$.  At this resolution
	the 2nd and 3rd order method look similar.  Both methods use 3rd order TVD Runge-Kutta time stepping.  Instabilities
	develop out of the fluid jets that emerge from the x-point when they collide with the slow moving fluids in the lobes.}\label{F:HighResReconnect}
\end{figure}
The simulations are run at two resolutions, $128\times 64$ and $512\times 256$ using a second or third order 
discontinuous Galerkin method with third
order TVD Runge Kutta time stepping.  At a resolution of $512\times 256$, 80 thousand time steps are taken
and at a resolution of $128\times 64$, 20 thousand time steps are taken.
In figure \ref{F:LowResReconnect} low resolution solutions to the GEM challenge problem are 
plotted at $t=25/\omega_{c\,i}$.  The second
order solution shows stable island formation while in the third order method no island forms.  
In this simulation the TVB limiter constant $M=0$
is used to eliminate the formation of an unstable island in the third order method.
Solutions to the GEM challenge problem are highly susceptible to bifurcation \cite{Hesse2001} which is 
the formation of magnetic islands
at or near the x-point (the center of the domain in this case).   The development of these islands
may be due to excessive dissipation applied at the x-point; however, the development of islands 
can be unpredictable and in the case of the 3rd order method, 
increased dissipation can actually eliminate the island. 

The islands are unstable and any small perturbation in the 
location or fields in the island will cause the island to slip and merge with one of the lobes.  
Small perturbations can arise from
machine precision error and this is particularly true at locations near equilibrium where two nearly equal but opposing 
forces are added, the number
that remains may depend significantly on the precision of the numbers used.  It has been observed that increasing
machine precision can change the direction that an unstable island slips off to.
The second order solution shows the formation of a stable island presumably
due to the extra dissipation of the second order method.

In figure \ref{F:HighResReconnect} the high resolution solutions are plotted at $t=25/\omega_{c\,i}$ with the TVB limiter 
constant $M=25$.  At this resolution the second and third order methods produce very similar results. 
In this case no island formation is visible in either the second or third order schemes, however at time 
$t=30/\omega_{c\,i}$ a very small
unstable island forms in the 3rd order method and combines with the left lobe before $t=35/\omega_{c\,i}$.  
\begin{figure}
	\begin{center}
	\scalebox{0.7}{\includegraphics{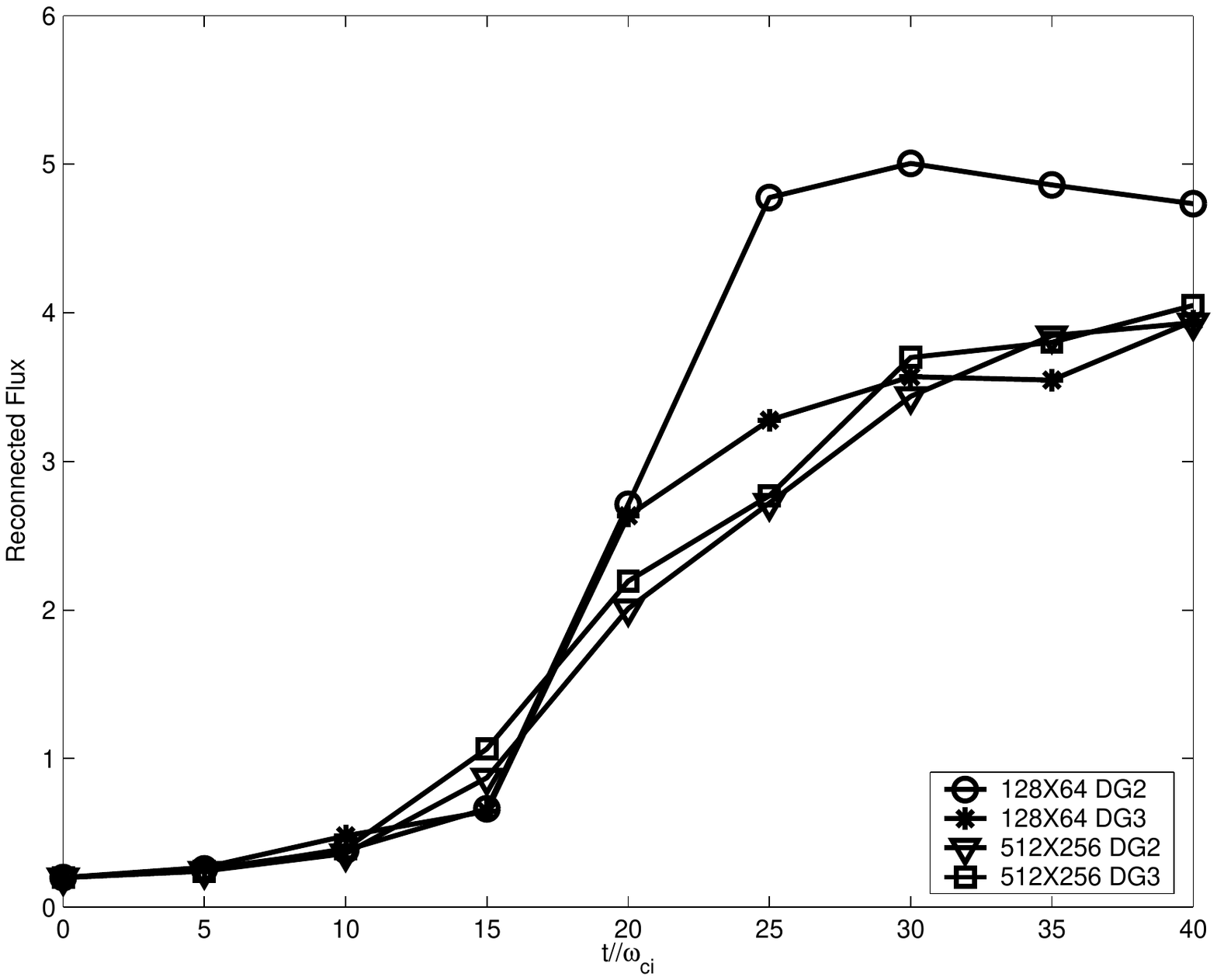}}
	\end{center}
	\caption{Plot of reconnected flux vs time for 2nd and 3rd order spatial discretizations at resolutions of
	$128\times 64$ and $512\times 256$.  The reconnected flux differs substantially for the two methods at $128\times 64$, 
	though the 3rd order method better matches the high order solutions.  At $512\times 128$ the 2nd and 3rd 
	order methods are in close agreement.}\label{F:CompareReconnectedFlux}
\end{figure}
In figure \ref{F:CompareReconnectedFlux} the reconnected magnetic flux $\frac{1}{2}\int|B_{y}|dx$ along the x axis 
is plotted for several
solutions.  At grid resolution of $512\times 256$ the second and third order methods produce similar results.  At a 
resolution of 
$128 \times 64$ the third order method produces results which are in much better agreement with the high 
resolution results than does the second order method.
%\begin{figure}
%	\begin{center}
%	\scalebox{1.5}{\includegraphics{figs/jcp-shay.pdf}}
%	\end{center}
%	\caption{Comparison of the two-fluid solution using the 3rd order discontinuous Galerkin method with
%	TVD 3rd order Runge-Kutta time stepping vs previously published solutions in \cite{Shay2001}.  This plot
%	was taken from \cite{Shay2001} and the two-fluid solution superimposed on top.  The two-fluid solution
%	shows good agreement with the other methods.  The Hall MHD code uses a term which depends on electron
%	mass and thus differs from other published Hall MHD solutions where the electron mass is taken
%	to be 0.}\label{F:ShayFlux}
%\end{figure}
%In figure \ref{F:ShayFlux} the solution using the 3rd order discontinuous Galerkin method with 3rd order TVD Runge-Kutta times
%stepping is compared with solutions produced with particle, hybrid and Hall MHD codes in \cite{Shay2001}.  The two-fluid solution
%agrees with all the converged solutions until near the point of saturation.  Physically, the reconnected solution must eventually 
%saturate.  

\begin{figure}
	\begin{center}
	\scalebox{0.7}{\includegraphics{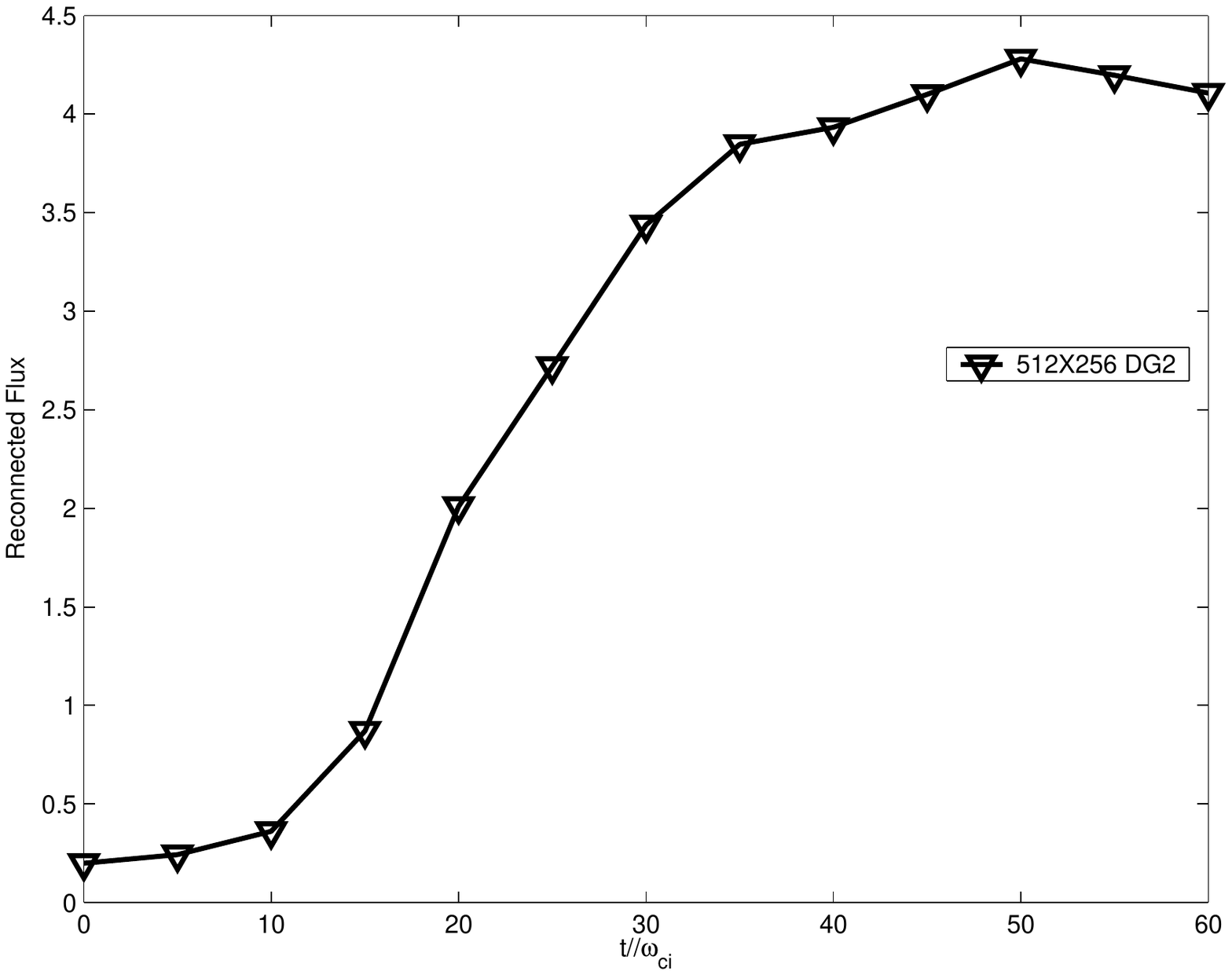}}
	\end{center}
	\caption{Plot of reconnected magnetic flux to time $t=60/\omega_{c\,i}$ for a $512X256$ solution.  This figure shows that
	the two-fluid solution does saturate as would be expected with conducting walls and finite flux in the domain.}\label{F:LongReconnect}
\end{figure}

In figure \ref{F:LongReconnect} the reconnected flux up to time $t=60/\omega_{c\,i}$ is shown illustrating the saturation
of the numerical solution beyond the published time interval in the GEM challenge results \cite{Birn2001}.

\begin{figure}
	\begin{center}
	\scalebox{0.7}{\includegraphics{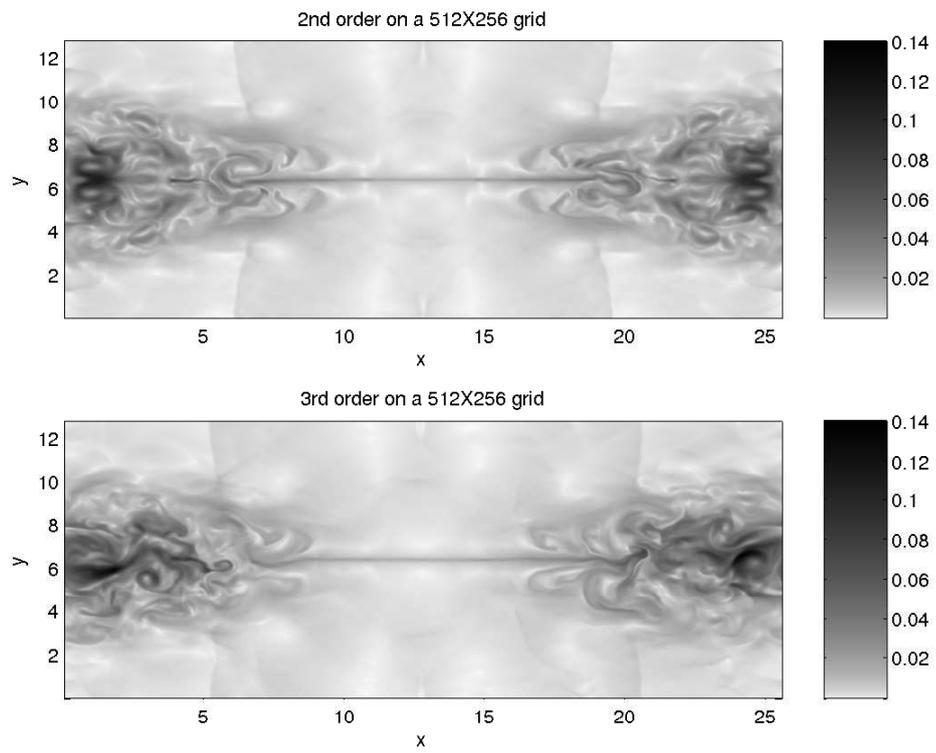}}
	\end{center}
	\caption{Plot of total ion momentum at $t=40/\omega_{c\,i}$ at a resolution 
	of $512\times 256$ using the second and third order DG methods.}
	\label{F:IonTurbulence}
\end{figure}

Flows appear late in time which are turbulent as shown in figure \ref{F:IonTurbulence} at time $t=40/\omega_{c\,i}$.
The 3rd order method shows considerably more asymmetry than the 2nd order method, but asymmetry has begun to develop in the 2nd order
solution.  This asymmetry develops out of machine precision errors that are initiated in the current layer at the x-point from the
balancing of source terms and fluxes.  The
only dissipation present is numerical so the second order method tends smooth out the errors produced by the finite precision more
than the 3rd order method.  As a result of the low dissipation in the 3rd order method these errors result 
in the excitation of unstable modes which eventually effect the macroscopic solution.  The magnitude of electron momentum
can differ by as much as 10 percent by moving from 64 bit numbers to 80 bit numbers at $t=40/\omega_{c\,i}$; fortunately
these regions tend to be localized.

Notice the pair of shocks in figure \ref{F:IonTurbulence}, the positions differ in the two solutions.  The positions differ because the onset
of the fast growth stage is slightly different for the two methods.  As the fast growth begins, shocks form in the ion fluid as it is accelerated
along the $x$ axis.  Eventually the two jets of shocked ion fluid collided and two shock that spans the $y$ axis are formed which continues to propagate
through the domain as the solution evolves.  If the onset of fast growth differs slightly initially, then these shocks will appear at different
locations for a given time.
\clearpage

\section{Discussion}\label{S:Discussion}
A discontinuous Galerkin method for the ideal 5-moment two-fluid plasma system is developed. 
A scalar model problem of the ideal two-fluid system is
derived to illustrate the character of the full system.  
An analytic two-fluid solution to an electron acoustic square pulse in the linear regime is derived
and the numerical solution using the fully non-linear two-fluid system is calculated showing convergence
for the 2nd and 3rd order discontinuous Galerkin methods.  The algorithm is benchmarked against the 
two-fluid electromagnetic shock originally published in \cite{Shumlak2003}. The 2nd and 3rd order algorithms 
are tested on the GEM challenge magnetic 
reconnection problem and produces results comparable to those generated by particle codes, hybrid codes, and a 
Hall MHD code in \cite{Shay2001}.  The discontinuous Galerkin method offers a straight forward method for solving the ideal 5-moment two-fluid system.  This same algorithm could be applied to 10 and higher moment two-fluid systems.  The algorithm can 
be easily generalized to three dimensions and general geometries.
\section*{Acknowledgments}
This work was supported by AFOSR Grant No. F49620-02-1-0129
\bibliographystyle{elsart-num}
\bibliography{dissertation}

\begin{thebibliography}{10}
\expandafter\ifx\csname url\endcsname\relax
  \def\url#1{\texttt{#1}}\fi
\expandafter\ifx\csname urlprefix\endcsname\relax\def\urlprefix{URL }\fi

\bibitem{Shumlak2003}
U.~Shumlak, J.~Loverich, Approximate riemann solver for the two-fluid plasma
  model, Journal of Computational Physics 187 (2003) 620--638.

\bibitem{Birn2001}
J.~Birn, et~al., Geospace environmental modeling (gem) magnetic reconnection
  challenge, Journal of Geophysical Research 106~(A3) (2001) 3715--3719.

\bibitem{Freidberg87}
J.~P. Freidberg, Ideal Magnetohydrodynamics, Plenum Press, 1987.

\bibitem{Bellan2000}
P.~M. Bellan, Spheromaks, Imperial College Press, 2000.

\bibitem{Hakim2007}
A.~Hakim, U.~Shumlak, Two-fluid physics and field-reversed configurations,
  Physics of Plasmas 14~(5) (2007) 055911.

\bibitem{Connor93}
J.~W. Connor, Pressure gradient turbulent transport and collisionless
  reconnection, Plasma Physics and Controlled Fusions 35 (1993) 757--763.

\bibitem{Krall71}
N.~A. Krall, P.~C. Liewer, Low-frequency instabilities in magnetic pulses,
  Physical Review A 4~(5) (1971) 2094--2103.

\bibitem{Davidson75}
R.~C. Davidson, N.~T. Gladd, Anomalous transport properties associated with the
  lower-hybrid-drift instability, The Physics of Fluids 18~(10) (1975)
  1327--1335.

\bibitem{Loverich2006}
J.~Loverich, U.~Shumlak, Nonlinear full two-fluid study of m = 0 sausage
  instabilities in an axisymmetric z pinch, Physics of Plasmas 13~(8) (2006)
  082310.

\bibitem{Solovev84}
L.~Solovev, Dynamics of a cylindrical z pinch, Soviet Journal of Plasma Physics
  10~(5) (1984) 602--605.

\bibitem{Haines82}
M.~Haines, The physics of the dense z-pinch in theory and in experiment with
  application to fusion reactor, Physica Scripta T2/2 (1982) 380--390.

\bibitem{Eberhardt96}
O.~S. Jones, U.~Shumlak, D.~S. Eberhardt, An implicit scheme for nonideal
  magnetohydrodynamics, Journal Of Computational Physics 130 (1997) 231--242.

\bibitem{Sovinec2004}
C.~Sovinec, et~al., Nonlinear magnetohydrodynamics simulation using high-order
  finite elements, Journal of Computational Physics 195 (2004) 355--386.

\bibitem{Bhattacharjee2003}
A.~Bhattacharjee, Center for magnetic reconnection studies: Present status,
  future plans, PSACI PAC Presentation, Princeton (June 2003).

\bibitem{Breslau2003}
J.~Breslau, S.~Jardin, A parallel algorithm for global magnetic reconnection
  studies, Computer Physics Communications 151 (2003) 8--24.

\bibitem{Huba2003}
J.~D. Huba, Hall magnetohydrodynamics - a tutorial, in: M.~S. J.~Buchner,
  C.T.~Dunn (Ed.), Space Plasma Simulation, Springer, 2003, pp. 166--192.

\bibitem{Park99}
W.~Park, et~al., Plasma simulation studies using multilevel physics models,
  Physics of Plasmas 6~(5) (1999) 1796--1803.

\bibitem{Baboolal2001}
S.~Baboolal, Finite-difference modeling of solitons induced by a density hump
  in a plasma multi-fluid, Mathematics and Computers in Simulation 55 (2001)
  309--316.

\bibitem{Munz95}
R.~Schneider, C.~D. Munz, The approximation of two-fluid plasma flow with
  explicit upwind schemes, International Journal of Numerical Modelling:
  Electronic Networks, Devices and Fields 8 (1995) 399--416.

\bibitem{Rambo91}
P.~Rambo, J.~Denavit, Time-implicit fluid simulation of collisional plasmas,
  Journal Of Computational Physics 98 (1991) 317--331.

\bibitem{Mason87}
R.~Mason, An electromagnetic field algorithm for 2d implicit plasma simulation,
  Journal of Computational Physics 71 (1987) 429--473.

\bibitem{Mason86}
R.~Mason, C.~Cranfill, Hybrid two-dimensional electron transport in
  self-consistent electromagnetic fields, IEEE Transactions on Plasma Science
  14~(1) (1986) 45--52.

\bibitem{Hakim2006}
A.~Hakim, J.~Loverich, U.~Shumlak, A high resolution wave propagation scheme
  for ideal two-fluid plasma equations, J. Comput. Phys. 219~(1) (2006)
  418--442.

\bibitem{Cockburn89}
B.~Cockburn, C.-W. Shu, Tvb runge-kutta local projection discontinuous galerkin
  finite element method for conservation laws ii: General framework,
  Mathematics of Computation 52 (1989) 411--435.

\bibitem{Cockburn89b}
B.~Cockburn, S.-Y. Lin, C.-W. Shu, Tvb runge-kutta local projection
  discontinuous galerkin finite element method for conservation laws iii:
  One-dimensional systems, Journal of Computational Physics 84 (1989) 90--113.

\bibitem{Cockburn90}
B.~Cockburn, S.~Hou, C.-W. Shu, The runge-kutta local projection discontinuous
  galerkin finite element method for conservation laws iv: Multidimensional
  case, Mathematics of Computation 54 (1990) 545--581.

\bibitem{Cockburn98}
B.~Cockburn, C.-W. Shu, The runge-kutta discontinuous galerkin method for
  conservation laws v: Multidimensional systems, Journal of Computational
  Physics 141 (1998) 199--224.

\bibitem{Hesthaven2002}
J.~S. Hesthaven, T.~Warburton, Nodal high-order methods on unstructured grids,
  Journal of Computational Physics 181 (2002) 186--221.

\bibitem{Cockburn2004}
B.~Cockburn, F.~Li, C.-W. Shu, Locally divergence-free discontinuous galerkin
  methods for maxwell's equations, Journal of Computational Physics 194 (2004)
  588--610.

\bibitem{Warburton99}
T.~C. Warburton, G.~E. Karniadakis, A discontinuous galerkin method for the
  viscous mhd equations, Journal of Computational Physics 152 (1999) 608--641.

\bibitem{Lin2006}
G.~Lin, G.~Karniadakis, A discontinuous galerkin method for two-temperature
  plasmas, Comput. Methods Appl. Mech. Engrg. 195 (2006) 3504--3527.

\bibitem{Mangeney2002}
A.~Mangeney, F.~Califano, C.~Cavazzoni, P.~Travnicek, A numerical scheme for
  the integration of the vlasov-maxwell system of equations, Journal of
  Computational Physics 179 (2002) 495--538.

\bibitem{Cockburn2000}
B.~Cockburn, G.~E. Karniadakis, C.-W. Shu (Eds.), Discontinuous Galerkin
  Methods, Springer, 2000.

\bibitem{Munz2000}
C.~D. Munz, R.~Schneider, U.~Vos, A finite-volume method for the maxwell
  equations in the time domain, Siam Journal of Scientific Computing 22 (2000)
  449--475.

\bibitem{Umeda2003}
T.~{Umeda}, Y.~{Omura}, T.~{Tominaga}, H.~{Matsumoto}, {A new charge
  conservation method in electromagnetic particle-in-cell simulations},
  Computer Physics Communications 156 (2003) 73--85.

\bibitem{Villasenor1992}
J.~Villasenor, O.~Buneman, Rigorous charge conservation for local
  electromagnetic field solvers, Computer Physics Communications 69~(2-3)
  (1992) 306--316.

\bibitem{Mardahl1997}
P.~J. {Mardahl}, J.~P. {Verboncoeur}, {Charge conservation in electromagnetic
  PIC codes; spectral comparison of Boris/DADI and Langdon-Marder methods},
  Computer Physics Communications 106 (1997) 219--229.

\bibitem{Li2008}
S.~Li, High order central scheme on overlapping cells for magneto-hydrodynamic
  flows with and without constrained transport method, J. Comput. Phys.
  227~(15) (2008) 7368--7393.

\bibitem{Balsara2009}
D.~S. Balsara, Divergence-free reconstruction of magnetic fields and weno
  schemes for magnetohydrodynamics, J. Comput. Phys. 228~(14) (2009)
  5040--5056.

\bibitem{Gardiner2008}
T.~A. Gardiner, J.~M. Stone, An unsplit godunov method for ideal mhd via
  constrained transport in three dimensions, J. Comput. Phys. 227~(8) (2008)
  4123--4141.

\bibitem{Londrillo2004}
P.~Londrillo, L.~D. Zanna, On the divergence-free condition in godunov-type
  schemes for ideal magnetohydrodynamics: the upwind constrained transport
  method, J. Comput. Phys. 195~(1) (2004) 17--48.

\bibitem{Durran98}
D.~Durran, Numerical Methods for Wave Equations in Geophysical Fluid Dynamics,
  Springer, 1998.

\bibitem{Leveque2002}
R.~J. LeVeque, Finite Volume Methods for Hyperbolic Problems, Cambridge
  University Press, 2002.

\bibitem{Biswas94}
R.~Biswas, K.~D. Devine, J.~E. Flaherty, Parallel, adaptive finite element
  methods for conservation laws, Applied Numerical Mathematics 14 (1994)
  255--283.

\bibitem{Brio88}
M.~Brio, C.~C. Wu, An upwind differencing scheme for the equations of ideal
  magnetohydrodynamics, Journal Of Computational Physics 75 (1988) 400--422.

\bibitem{Loverich2003}
J.~Loverich, A finite volume algorithm for the two-fluid plasma system in one
  dimension, Masters thesis, University of Washington (2003).

\bibitem{Loverich2005}
J.~Loverich, U.~Shumlak, A discontinuous galerkin method for the full two-fluid
  plasma model, Computer Physics Communications 169 (2005) 251--255.

\bibitem{Hesse2001}
M.~Hesse, J.~Birn, M.~Kusnetsova, Collisionless magnetic reconnection: Electron
  processes and transport modeling, Journal of Geophysical Research 106 (2001)
  3721--3735.

\bibitem{Shay2001}
M.~A. Shay, J.~F. Drake, B.~N. Rogers, R.~E. Denton, Alfvenic collisionless
  magnetic reconnection and the hall term, Journal of Geophysical Research 106
  (2001) 3759--3772.

\end{thebibliography}
\end{document}